\documentclass[journal, web]{IEEEtran}
\usepackage{cite}
\usepackage{amsmath,amssymb,amsfonts}
\usepackage{algorithmic}
\usepackage{subfig}
\usepackage{graphicx}
\usepackage{textcomp}
\usepackage{wrapfig}
\usepackage{tabularx}
\usepackage{hyperref}
\usepackage{mathtools}
\usepackage{soul}
\usepackage[dvipsnames]{xcolor}

\def\BibTeX{{\rm B\kern-.05em{\sc i\kern-.025em b}\kern-.08em
    T\kern-.1667em\lower.7ex\hbox{E}\kern-.125emX}}
\begin{document}
\title{{DoorINet}: Door Heading Prediction through Inertial Deep Learning}
\author{Aleksei Zakharchenko, Sharon Farber, Itzik~Klein
% \thanks{Manuscript received August 4, 2024.}
%\thanks{Aleksei Zakharchenko is supported by the Maurice Hatter Foundation}
\thanks{Aleksei Zakharchenko, Sharon Farber and Itzik Klein are with The Autonomous Navigation and Sensor Fusion Lab, Hatter Department of Marine Technologies, University of Haifa, Israel. Corresponding author: azakharc@campus.haifa.ac.il.}
  }
\maketitle

\begin{abstract}
Inertial sensors are widely used in a variety of applications. A common task is orientation estimation. To tackle such a task, attitude and heading reference system algorithms are applied. Relying on the gyroscope readings, the accelerometer measurements are used to update the attitude angles, and magnetometer measurements are utilized to update the heading angle. In indoor environments, magnetometers suffer from interference that degrades their performance resulting in poor heading angle estimation. Therefore, applications that estimate the heading angle of moving objects, such as walking pedestrians, closets, and refrigerators, are prone to error. To circumvent such situations, we propose DoorINet, an end-to-end deep-learning framework to calculate the heading angle from door-mounted, low-cost inertial sensors without using magnetometers. To evaluate our approach, we record a unique dataset containing 391 minutes of accelerometer and gyroscope measurements and corresponding  ground-truth heading angle. We show that our proposed approach outperforms commonly used, model based approaches and data-driven methods. To enable reproducibility of our results and future research, both code and data are available at \url{https://github.com/ansfl/DoorINet}.
\end{abstract}


\section{Introduction}
\label{sec:introduction}
\IEEEPARstart{I}{nertial} sensors are found in many applications such as motion capture and human activity recognition \cite{6365160,sensors_activity_rec,9261582,har_alg_imu1}, sports applications \cite{electronics8111257,sensors_sports,WU2022100494,GHOSH2022101608}, health-related applications \cite{Qadri2020TheFO,hiot_emerging,health_sensors}, pedestrian dead reckoning \cite{  8231129,9090160,8960327}, and smart home applications \cite{smart_home_survey,review_iot,smart_home_design}. \\
Inertial measurement units (IMU) consist of three mutually orthogonal accelerometers and three gyroscopes, aligned with the accelerometers. Additionally, there can be three orthogonal magnetometers, aligned in the same direction as other sensors. Accelerometers provide the specific force vector, gyroscopes - the angular rate vector, and the magnetometers - the magnetic field vector. IMUs come in a wide range of performance grades intended for different applications. Though there is no universally agreed definition of high-, medium-, and low-grade IMUs, they can be grouped into several categories: marine, aviation, intermediate, tactical, and automotive \cite{9101092}. Automotive grade is sometimes called consumer or commercial grade. The most affordable IMU sensors are usually built with micro electro-mechanical sensor (MEMS) technology, also making them small compared to mechanical IMUs. Miniaturized MEMS IMUs are widely used in consumer products, i.e., smartphones, IoT applications, and wearable devices. \\
In non-navigational applications (such as most IoT applications), inertial sensor readings are used as input to attitude and heading reference system (AHRS) algorithms that provide the orientation of a device; that is, the three Euler angles with respect to a fixed coordinate frame, which uniquely determine the orientation.  \\
In model-based approaches, the AHRS algorithm divides into two separate independent parts:  orientation propagation from a gyroscope, and readings followed by updates from accelerometer and magnetometer measurements. In nonlinear filtering approaches the attitude kinematic equations are treated as the system model, and the accelerometer and magnetometer observations are processed in the measurement model to update the state and propagate the error-state covariance matrix \cite{aided_navigation_farrell_book}. Another commonly used approach is complementary filtering, which combines compensated gyroscope measurements with  gain-multiplied measurements from accelerometers and magnetometers \cite{madgwphd, mahony_algo} \\
In learning-based approaches, IMU readings are fed directly into the deep-learning algorithm to provide attitude and heading estimation \cite{ASGHARPOORGOLROUDBARI2023113105}. Alternatively, a deep-learning algorithm can be used for denoising inertial measurements, and attitude estimation is obtained using an integrated processed "noise-free" signal like in \cite{lgcnet}. \\
Hybrid approaches try to combine the best of model and learning approaches. For example, the deep attitude estimator \cite{9333630,EranDAE, Vertzberger2021} uses a deep-learning model to calculate gain for the complementary filter algorithm structure. \\
% Accelerometer and gyroscope readings together are used to obtain attitude angles relative to the celestial horizon (a plane that is perpendicular to the gravity vector). The heading angle which corresponds to rotation around the axis parallel to the gravity vector is usually the most hard to calculate, and in most classical AHRS algorithms such calculation requires magnetometer measurements. \\
\noindent
The heading information is critical for several IoT applications such as smart homes, smart offices, or building management. It is also important for space planning, furniture placement and safety. The heading of a door can be estimated using magnetometers. Yet, indoor environments are full of electromagnetic fields, resulting in constant and time varying influence on magnetometer readings. This is even more relevant in specific industrial buildings with high levels of magnetic field, where the heading angle accuracy degrades. Consequently, magnetometers cannot be used to determine the door heading angle. To cope with this situation we propose DoorINet, a deep learning end-to-end framework for estimating the heading angle of door-mounted IMU using only accelerometers and gyroscopes. Two versions of DoorINet are examined: AG-DoorINet that takes accelerometer and gyroscope measurements, and G-DoorINet that takes only gyroscope readings.\\
\noindent
To evaluate our proposed approaches we rrecord a unique dataset using Xsens DOT IMUs mounted on three different doors, resulting in 391 minutes of recorded inertial data and accurate headings obtained from a higher grade Memsense IMU. We compare our approach with two model-based and three learning-based AHRS approaches, and show that ours outperform the rest on a real-life scenario of an inner door over 90 minutes. Our deep-learning framework is able to generalize over different error parameters of different IMU sensors, and it is able to correctly calculate the heading angle. \\
The contributions of this paper are as follows:
\begin{enumerate}
    \item Derivation of two end-to-end networks capable of accurately regressing the heading angle of a door-mounted inertial sensors using only accelerometer and gyroscope readings.
    \item Recording of a unique dataset containing 391 minutes of accelerometer and gyroscope measurements and corresponding  ground-truth (GT) heading angle.
    \item Ensuring reproducibility of our results and encouraging future research by making the code and dataset available at \href{https://github.com/ansfl/DoorINet}{this} GitHub repository.
\end{enumerate}
% We tested DoorINet in a real-life scenario of over 90 minutes long and we have shown that both versions of DoorINet generalize well over different sensor drift parameters and perform better than model-based AHRS, reducing sensor drift. That makes it sensible to use lower-grade MEMS Xsens DOT IMUs in applications that require working non-stop in the longer time period. 

\noindent
\noindent
The rest of the paper is organized as follows: Section \ref{sec:problem_formulation} introduces model-based approaches for heading estimation. Section \ref{sec:proposed_approach} describes our proposed learning approach for heading determination. Section \ref{sec:dataset_generation} explains our dataset collection setup and describes post-processing procedures. Section \ref{sec:results} shows the experimental results and Section \ref{sec:conclusions} concludes this work.

\section{Heading Estimation Approaches} \label{sec:problem_formulation}
%1807 as you dont use eq (1) and (2) in the paper this section is not relevant.
%Alex done

%1807 write here 2-3 sentences about that there are many different AHRS algo, yet the focus of the paper is only heading therefore we review a basic integration approach and a complementary filtering approach
%Alex done
\subsection{Model-based approaches}
\noindent
There are several model-based AHRS algorithms that estimate the attitude and heading of a platform. As this work focuses only on the heading angle, we briefly review the direct integration approach and the Madgwick complementary filtering approach.

\subsubsection{Direct Integration} \label{sec:gyro_form}
\noindent
Gyroscopes measure angular velocity, that is, the change of angle with respect to time. Assuming the gyroscope's sensitive axis is aligned with the gravity direction, the heading angle $\psi$ can be found by integrating the angular velocity $w$ component parallel to the gravity direction:
\begin{equation}\label{eq:gyro_integration}
    \psi(t) = \psi_0 + \int_{t}w_t\,dt \,
\end{equation}
where $\psi_0$ is the initial heading angle. This approach does not take into account error factors in the gyroscope sensor and alignment errors, and, therefore, is considered less accurate than AHRS methods.

\subsubsection{Madgwick Filter} \label{sec:madgwick_alg}
\noindent
Madgwick first proposed an orientation filter for inertial and inertial/magnetic sensor arrays in \cite{madgw_internal}, and later its extended version \cite{madgw}. We follow a version of the Madgwick algorithm as presented in chapter 7 of \cite{madgwphd}. This algorithm functions as a complementary filter that combines compensated gyroscope measurements with filtered measurements from the accelerometer with frequency determined by the gain. \\
The orientation is obtained from integration of the quaternion derivative describing the rate of change of orientation of the earth frame $E$ relative to the sensor frame $S$, by

\begin{equation}\label{eq:madgwick_orientation_integral}
\prescript{S}{E}{\boldsymbol{\mathrm{q}}} = \int_t \prescript{S}{E}{\Dot{\boldsymbol{\mathrm{q}}}}\,dt,
\end{equation}
where $\prescript{S}{E}{\boldsymbol{\mathrm{q}}_{\omega,t}}$ is the unnormalized orientation and $\prescript{S}{E}{\Dot{\boldsymbol{\mathrm{q}}}_{\omega,t}}$ is the quaternion derivative defined by
\begin{equation}\label{eq:madgwick_orientation_derivative}
\prescript{S}{E}{\Dot{\boldsymbol{q}}} = \frac{1}{2} \prescript{S}{E}{\hat{\boldsymbol{\mathrm{q}}}} \otimes  
\begin{matrix} 
{[ 0,  (\boldsymbol{\mathrm{\omega}}^{\prime} + Ke) ^ {T} ]}
% {\Bigl[]}
\end{matrix},
\end{equation}
where $\prescript{S}{E}{\hat{\boldsymbol{\mathrm{q}}}}$ is a normalized orientation in a quaternion form,  $\otimes$ is the quaternion product, $\boldsymbol{\mathrm{\omega}}^{\prime}$ is the gyroscope measurement vector, $K$ is the algorithm gain, and $e$ is an error term. \\
% \begin{equation} \label{eq:madgwick_omega_prime}
%     \omega^{\prime} = \omega - 2 \pi f_c \int p \omega \, dt,
% \end{equation}
% where $p$ is a value dynamically determined as 1 or 0 to enable integrating only when the IMU is stationary, and $f_c$ is a corner frequency.
The initial value of $\prescript{S}{E}{\hat{\boldsymbol{\mathrm{q}}}}$ is assumed to be an identity quaternion, and the algorithm to determine the gain K is

\begin{equation}\label{eq:madgwick_algorithm_gain}
K = \begin{cases}
    K_{norm} + \frac{t_{init} - t}{t_{init}} (K_{init} - K_{norm}), & \text{if } t<t_{init} \\
    K_{norm} & \text{else}
\end{cases}
\end{equation}
where $K_{init}$ is the initial large value of gain, $K_{norm}$ is the value intended for normal operation, and $t_{init}$ is the initialization period. \\
The error term used in \eqref{eq:madgwick_orientation_derivative} is defined by

\begin{equation} \label{eq:madgwick_error_factor}
e = \begin{cases}
    % e_f + e_m, & \text{if } ||\boldsymbol{\mathrm{f}}||>0 \text{ and } ||\boldsymbol{\mathrm{m}}|| > 0, \\
    e_f, & \text{else if } ||\boldsymbol{\mathrm{f}}|| > 0, \\
    \begin{bmatrix} 0 & 0 & 0  \end{bmatrix} ^T & \text{else},
\end{cases}
\end{equation}
where $\boldsymbol{\mathrm{f}}$ is the specific force measurement vector, and $e_f$ is the accelerometer error factor defined as

\begin{equation} \label{eq:madgwick_ea}
    \boldsymbol{e}_f = \hat{\boldsymbol{\mathrm{f}}} \times \begin{bmatrix}
        2 q_x q_y - 2 q_w q_z \\
        2 q^2_w - 1 + 2 q^2_y \\
        2 q_y q_z - 2 q_w q_x
    \end{bmatrix}, \hat{\boldsymbol{\mathrm{f}}} = \frac{\boldsymbol{\mathrm{f}}}{||\boldsymbol{\mathrm{f}}||} 
\end{equation}
% and 

% \begin{equation} \label{eq:madgwick_em}
%     e_m = \frac{\boldsymbol{\mathrm{f}} \times \boldsymbol{\mathrm{m}}}{||\boldsymbol{\mathrm{f}} \times \boldsymbol{\mathrm{m}}||} \times \begin{bmatrix}
%         2 q_x q_z - 2 q_w q_y \\
%         2 q_y q_z - 2 q_w q_x \\
%         2 q^2_w - 1 + 2 q^2_z
%     \end{bmatrix}
% \end{equation}
\noindent
where $\boldsymbol{\mathrm{f}}$ is the specific force vector measured by the accelerometer, and $q_w$, $q_x$, $q_y$, $q_z$ are elements of $\prescript{S}{E}{\hat{\boldsymbol{\mathrm{q}}}}$. \\
Additional features of this algorithm are designed to address various challenges (sensor conditioning, gyroscope bias compensation, linear acceleration rejection, etc.) and are not described in this paper. %The complete algorithm is presented in \cite{madgwphd}.%

\subsection{Learning approaches}
\noindent
There exist a significant number of data-driven models for orientation estimation. Most of them solve a 3D estimation problem, which in our case is superficial as we are focused only on the heading angle estimation.  Three data-driven approaches were implemented for comparison with our proposed approach:
\noindent
\subsubsection{\textbf{Deep Attitude Estimator (DAE)} \cite{EranDAE}}  combines both deep learning and a model-based backbone. It uses accelerometers and gyroscopes as input. The underlying idea of this approach is to estimate gain for a complementary filter algorithm using a neural network.
\noindent
\subsubsection{\textbf{Quaternion Model A} \cite{ASGHARPOORGOLROUDBARI2023113105}} consists of convolution layers, bi-directional LSTM layers, max-pooling, and fully connected layers, and can work with inertial sensors with different sampling rates. The goal of the paper was to achieve a real-time accurate three-dimensional attitude estimation using accelerometer and gyroscope readings.
% The model was trained on publicly available datasets by authors and open for public access, and we were able to test it on our test dataset.
% Authors of said models designed them that way that models generalizes over different inertial sensors with different sampling rate, they also trained their models on publicly available dataset that have 3-dimensional movement and orientation change. 
% We have been able to test one pre-trained model out of three proposed by the authors of aforementioned article, namely Model A (in this paper this model is called "Quaternion Model A"). It has a hybrid structure: 1-dimensional convolution layers, one bi-directional LSTM layer, maxpooling and fully-connected layers. The model requires that input data to be organized as sequences of 3-axes inertial sensors data of length 100 (accelerometer data 3 x 100 and gyroscope data 3 x 100), and the model outputs the absolute orientation in quaternion form. The resulting orientation belongs to the time period in the middle of the input sequence, that is, after first 50 readings.
\noindent
\subsubsection{\textbf{LGC-Net} \cite{lgcnet}} performs denoising of IMU readings of low-cost, MEMS-based inertial sensors and contains special kinds of layers: depthwise separable convolution  \cite{chollet2017xception} and large kernel attention \cite{guo2022visual} layers. The authors aimed to obtain clean gyroscope readings free from noise and bias, so the orientation could be obtained by directly integrating it. \\
We implemented the LGC-Net model based on its description in \cite{lgcnet} and trained it on our dataset. We used the \texttt{HuberLoss} loss function instead of the \texttt{Log-cosh} loss function implemented in the \cite{lgcnet}.

\section {Proposed Learning Approach for Heading Estimation} \label{sec:proposed_approach}
\begin{figure}[ht]
  \begin{center}
  \includegraphics[width=3.5in]{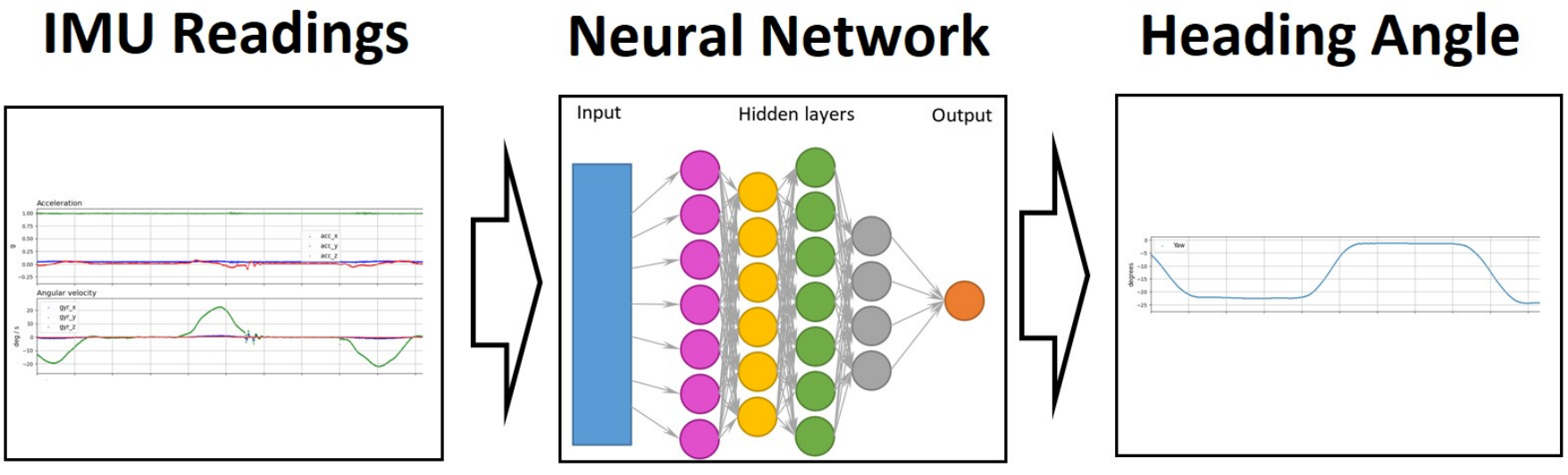}
  \caption{Our proposed data-driven approach.}\label{fig:flowchart}
  \end{center}
\end{figure}

%1807  start with what problem are you solving and what are the current algo' shortcomings. Then you need to state in general what is our solution - figure 1 - describe figure 1 in the text and then justify why NN. also lets work only with heading angle and not yaw.
%Alex done
\noindent
When working with IoT applications it is crucial to ensure the heading accuracy in longer time durations. Current AHRS algorithms either require fine-tuning to the current scenario to provide valid results, or are simply unable to provide them as low-cost inertial sensors are employed. \\
In this work we propose DoorINet, a deep-learning inertial framework for door-mounted IoT applications. To that end, we design a deep-learning architecture fed by accelerometer and gyroscope readings to regress the heading angle as presented in Figure \ref{fig:flowchart}. The advantages of our approach are twofold:

\begin{enumerate}
    \item No fine-tuning or weight parameters are required (in contrast to model-based approaches)
    \item Using deep-learning algorithms allows leveraging its well-known properties like noise reduction, the ability to cope with nonlinearities in the data, and the potential ability to generalize over different sensor error parameters.
\end{enumerate}
\vspace{-4mm}
\vspace{-4mm}
\noindent
As IoT applications are considered, we focused our design on relatively shallow architectures allowing their implementation for real-time settings. Two types of networks are proposed, differing in their structure and input, as addressed in the next section.
% Several neural network architectures were tested in order to determine which one provides the best results. The most promising were networks from Recurrent Neural Networks (RNN) family, which consist of RNN layers.
To find the most suitable network for heading estimation, we examined many different network architectures, including convolutional neural networks (CNN), long-short-term memory layers (LSTM), and gated recurrent networks (GRU). Of all these architectures, GRU was found the most promising and is described in the following sections.
\vspace{-4mm}
\subsection{Gyroscope-Only DoorINet Architecture (G-DoorINet)}
% I propose to form a subsection about all the RNNs in common, and next subsections about specifically our gyro-only and gyro+acc networks 
\noindent
Our network comprises bidirectional gated recurrent units (BiGRU)---a type of recurrent layers---and fully connected (FC) layers. A recurrent layer consists of a hidden state $\boldsymbol{\mathrm{h}}$ and an optional output $\boldsymbol{y}$ that operates on a variable-length sequence $\mathrm{\boldsymbol{x}} = (x_1, ... , x_T)$. At each time step \textit{t}, the hidden state $\boldsymbol{\mathrm{h}}_{\langle t \rangle}$ of the layer is updated by
\begin{equation}\label{eq:rnn_update_hidden}
\boldsymbol{\mathrm{h}}_{\langle t \rangle} = f(\boldsymbol{\mathrm{h}}_{\langle t-1 \rangle}, \boldsymbol{\mathrm{x}}_t),
\end{equation}
where $f$ is a non-linear activation function. \\
Bidirectional recurrent layers were introduced in \cite{BiRNN}, and they connect hidden states of two recurrent layers processing an input sequence in both forward and backward directions to the same output. \\
GRU layers are a specific type of RNN layer that was first suggested in \cite{cho2014learning} for natural language processing applications. GRU layers have two main gates allowing them to adaptively remember and forget information. The $\textit{reset}$ gate $r_j$ is computed by
\begin{equation}\label{eq:gru_reset_gate}
r_j = \sigma ( [\boldsymbol{\mathrm{W}}_r \boldsymbol{\mathrm{x}}]_j + [\boldsymbol{\mathrm{U}}_r \boldsymbol{\mathrm{h}}_{\langle t-1 \rangle}]_j),
\end{equation}
where $\sigma$ is the logistic sigmoid function, and $[.]_j$ denotes the $j$-th element of a vector. $\boldsymbol{\mathrm{x}}$ and $\boldsymbol{\mathrm{h}}_{t-1}$ are the input and the previous hidden state, respectively. $\boldsymbol{\mathrm{W}}_r$ and $\boldsymbol{\mathrm{U}}_r$ are weight matrices that are learned.  \\
The $\textit{update}$ gate $z_j$ is computed by

\begin{equation}\label{eq:gru_update_gate}
z_j = \sigma ( [\boldsymbol{\mathrm{W}}_z \boldsymbol{\mathrm{x}}]_j + [\boldsymbol{\mathrm{U}}_z \boldsymbol{\mathrm{h}}_{\langle t-1 \rangle}]_j).
\end{equation} 
\noindent
The actual activation of the proposed unit $h_j$ is then computed by

\begin{equation}\label{eq:gru_activation1}
 h_j^{\langle t \rangle} = z_j h_j^{\langle t-1 \rangle} + (1 - z_j) \Tilde{h}_j^{\langle t \rangle}, 
\end{equation}
where

\begin{equation}\label{eq:activation2}
\Tilde{h}_j^{\langle t \rangle} = \phi  ( [\boldsymbol{\mathrm{Wx}}]_j + [\boldsymbol{U} (\boldsymbol{\mathrm{r}}  \odot  \boldsymbol{\mathrm{h}}_{\langle t-1 \rangle}] )
\end{equation}
where $\odot$ is an element-wise (Hadamard) product.  \\
When the reset gate is close to zero, the hidden state is forced to ignore the previous hidden state and is reset with the current input only. This allows the hidden state to drop any information that is found to be irrelevant later in the future, allowing a more compact representation. On the other hand, the update gate controls how much information from the previous hidden state carries over to the current hidden state. \\
% This acts similarly to the memory cell in the LSTM network and helps the network to remember long term information.%1807 then add the following subsection A) Gyro based Network Structure B) Accl and Gyro Network Structure. you need to justify why you work with different network for each case.
%Alex done
\noindent
In addition to BiGRU layers, we employ fully connected (FC) layers, which were first introduced in \cite{rosenblatt_perc}, and calculate vector-matrix multiplication $\boldsymbol{\mathrm{x}} \times \boldsymbol{\mathrm{W}}$, where $\boldsymbol{\mathrm{x}}$ is the input vector and $\boldsymbol{\mathrm{W}}$ is the weight matrix. An activation function \texttt{tanh} is applied between fully connected layers element-wise and defined as

\begin{equation}
    tanh(\boldsymbol{\mathrm{x}}_j) = \frac{exp(\boldsymbol{\mathrm{x}}_j) - exp(-\boldsymbol{\mathrm{x}}_j)}{exp(\boldsymbol{\mathrm{x}}_j) - exp(-\boldsymbol{\mathrm{x}}_j)}
\end{equation}
where $\boldsymbol{\mathrm{x}}_j$ is a $j$-th element of a $tanh$ input. \\
% The proposed DoorINet neural network (our model) structure consists of GRU layers and fully-connected (FC) layers. We consider two variations of our model, the one that takes 3-axis gyroscope readings only and the other that takes 3-axis gyroscope and 3-axis accelerometer input. Since neural networks take input of a fixed size, we need two separate neural networks.
% \subsection{Gyroscope based DoorINet network structure}
\noindent
Figure \ref{fig:rnn_arch_3} describes DoorINet (G-DoorINet) architecture for the learning the door heading angle. It takes gyroscope readings as its input and fed them into the bi-directional GRU layer (marked as BiGRU1 box), which output then goes to another bi-directional GRU layer (BiGRU2). Next, the output of BiGRU2 goes to a sequence of eight fully connected layers, which outputs the heading angle:

\begin{figure}[!ht]
  \begin{center}
  \includegraphics[width=3.5in]{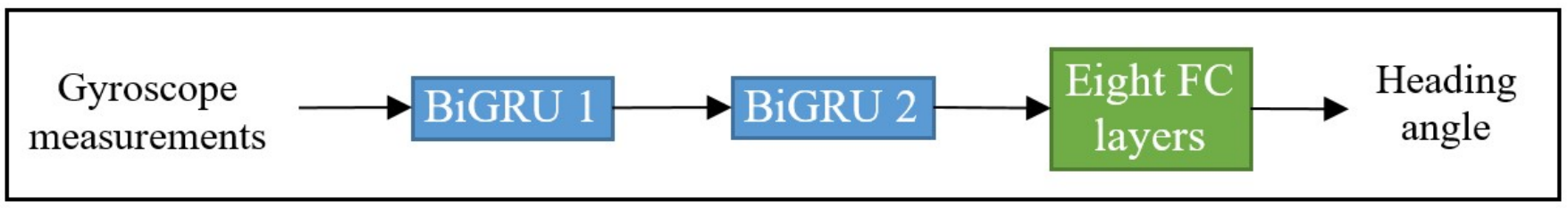}
  \caption{G-DoorINet network architecture using only gyroscope measurements.}\label{fig:rnn_arch_3}
  \end{center}
\end{figure} 
% \vspace{-4mm}
\noindent
The BiGRU layer that processes gyroscope readings (BiGRU1) accumulates knowledge about every sample (which is the sequence of 20 consequent readings) in one block of data. Because each ground-truth value is the angle difference between the first and the last heading angle of the IMU, the direction of data is not relevant, therefore use of bidirectional option of GRU layers is justified. The BiGRU2 deals with processed data from gyroscope. Fully connected layers are organized in common pyramid-like encoder fashion and serve as a function approximator that outputs one value for the series of readings.

\noindent
In Table \ref{tab:3axis_structure} we give G-DoorINet network parameters including input and output sizes.

\begin{table}[!ht] 
\caption{G-DoorINet network parameters.}
\begin{tabular}{cccc}
\multicolumn{1}{m{1.5cm}}{\centering \textbf{Layer}} & \textbf{Input size} & \textbf{Output size} & \multicolumn{1}{m{3cm}}{\centering \textbf{Additional}}   \\\\ \hline
\centering BiGRU 1 & 3          & 64          & \begin{tabular}[c]{@{}l@{}}bidirectional\end{tabular} \\ \hline
\centering BiGRU 2 & 128        & 128         & \begin{tabular}[c]{@{}l@{}}bidirectional\end{tabular}  \\ \hline
\centering FC 1  & 5120      & 2560        & \begin{tabular}[c]{@{}l@{}}Dropout p=0.2; Tanh\end{tabular} \\ \hline
FC 2  & 2560       & 512         & \begin{tabular}[c]{@{}l@{}}Tanh\end{tabular}             \\ \hline
FC 3  & 512        & 128         & Tanh                                                                       \\ \hline
FC 4  & 128        & 32          & Tanh                                                                       \\ \hline
FC 5  & 32         & 16          & Tanh                                                                       \\ \hline
FC 6  & 16         & 8           & Tanh                                                                       \\ \hline
FC 7  & 8          & 4           & Tanh                                                                       \\ \hline
FC 8  & 4          & 1           & -     \\ \hline
\end{tabular}
\label{tab:3axis_structure}
\end{table}
\vspace{-4mm}
\subsection{Accelerometer and Gyroscope  DoorINet Architecture (AG-DoorINet)}
\begin{figure}[ht!]
  \begin{center}
  \includegraphics[width=3.5in]{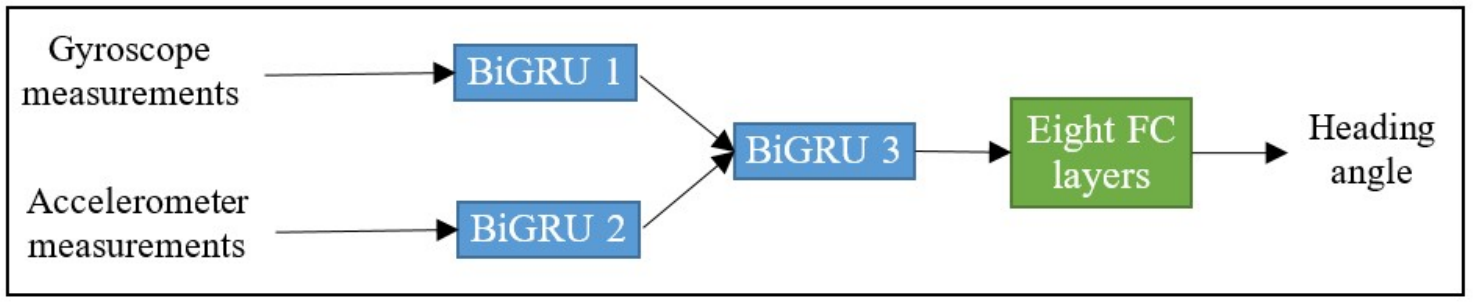}
  \caption{AG-DoorINet network architecture using accelerometer and gyroscope measurements.}\label{fig:rnn_arch_6}
  \end{center}
\end{figure}

\noindent
The accelerometer and gyroscope model (AG-DoorINet) consists of two sets of stacked, bi-directional GRU (BiGRU) layers that feed on accelerometer and gyroscope data (making it a multi-head neural network), whose outputs are fed to another GRU layer and FC layers afterwards, as illustrated in Figure  \ref{fig:rnn_arch_6}).

\noindent
The BiGRU layers that independently process accelerometer and gyroscope readings (BiGRU1 and BiGRU2) accumulate knowledge about every sample (which is the sequence of 20 consequent readings) in one block of data. Because each ground-truth value is the angle difference between the first and the last heading angle of the IMU, the direction of data is not relevant, therefore use of bidirectional option of GRU layers is justified. The BiGRU3 deals with the stacked processed data from accelerometer and gyroscope, taking it to the next logical layer. Fully connected layers are organized in common pyramid-like encoder fashion and serve as a function approximator that outputs one value for the series of readings.

\noindent
% \vspace{-4mm}
Table \ref{tab:6axis_structure} presents the AG-DoorINet network parameters including input and output sizes.
\begin{table}[ht!] 
\caption{AG-DoorINet network parameters.}
\begin{tabular}{cccc}
\multicolumn{1}{m{1.5cm}}{\centering \textbf{Layer}} & \textbf{Input size} & \textbf{Output size} & \multicolumn{1}{m{3cm}}{\centering \textbf{Additional}}   \\\\ \hline
BiGRU 1 & 3          & 64          & \begin{tabular}[c]{@{}c@{}}Bidirectional; \\ stacked layers=3\end{tabular} \\ \hline
BiGRU 2 & 3          & 64          & \begin{tabular}[c]{@{}c@{}}Bidirectional;\\ stacked layers=3\end{tabular}  \\ \hline
BiGRU 3 & 256        & 256         & \begin{tabular}[c]{@{}c@{}}Bidirectional;\\ stacked layers=2\end{tabular}  \\ \hline
FC 1  & 10240      & 2560        & \begin{tabular}[c]{@{}l@{}}Dropout p=0.2; Tanh\end{tabular}              \\ \hline
FC 2  & 2560       & 512         & \begin{tabular}[c]{@{}l@{}}Dropout p=0.2;  Tanh\end{tabular}             \\ \hline
FC 3  & 512        & 128         & Tanh                                                                       \\ \hline
FC 4  & 128        & 32          & Tanh                                                                       \\ \hline
FC 5  & 32         & 16          & Tanh                                                                       \\ \hline
FC 6  & 16         & 8           & Tanh                                                                       \\ \hline
FC 7  & 8          & 4           & Tanh                                                                       \\ \hline
FC 8  & 4          & 1           & -     \\ \hline
\end{tabular}
\label{tab:6axis_structure}
\end{table}

% \subsubsection{OriNet: Robust 3-D Orientation Estimation With a Single Particular IMU \cite{OriNet}}

% OriNet is a complex framework that uses only angular velocity measurements as its input. The architecture of OriNet includes LSTM and fully-connected layers. The model can take data with different sampling rate and outputs an absolute 3-dimensional orientation in quaternion form, which requires also taking the previous orientation as an input.

% Unfortunately we were not able to acquire the OriNet model or source code to test it on our data, so no comparison has been made.

\section{Dataset generation and processing} \label{sec:dataset_generation}
\begin{figure}[ht!]
  \centering
  \subfloat[Setups 1 and 2]{\includegraphics[width=0.235\textwidth]{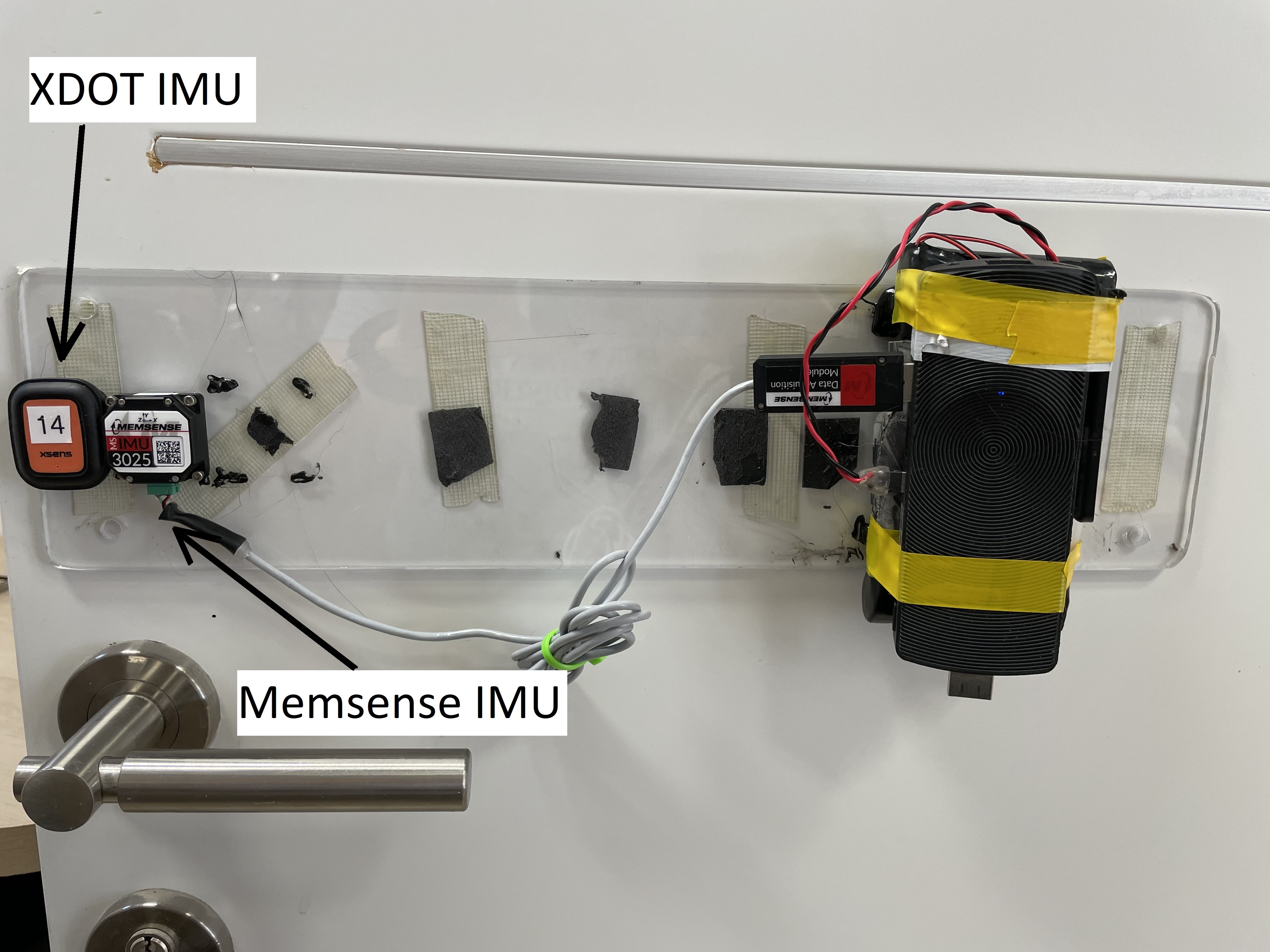}\label{fig:exp_1}}
  \hfill
  \subfloat[Setup 3]{\includegraphics[width=0.235\textwidth]{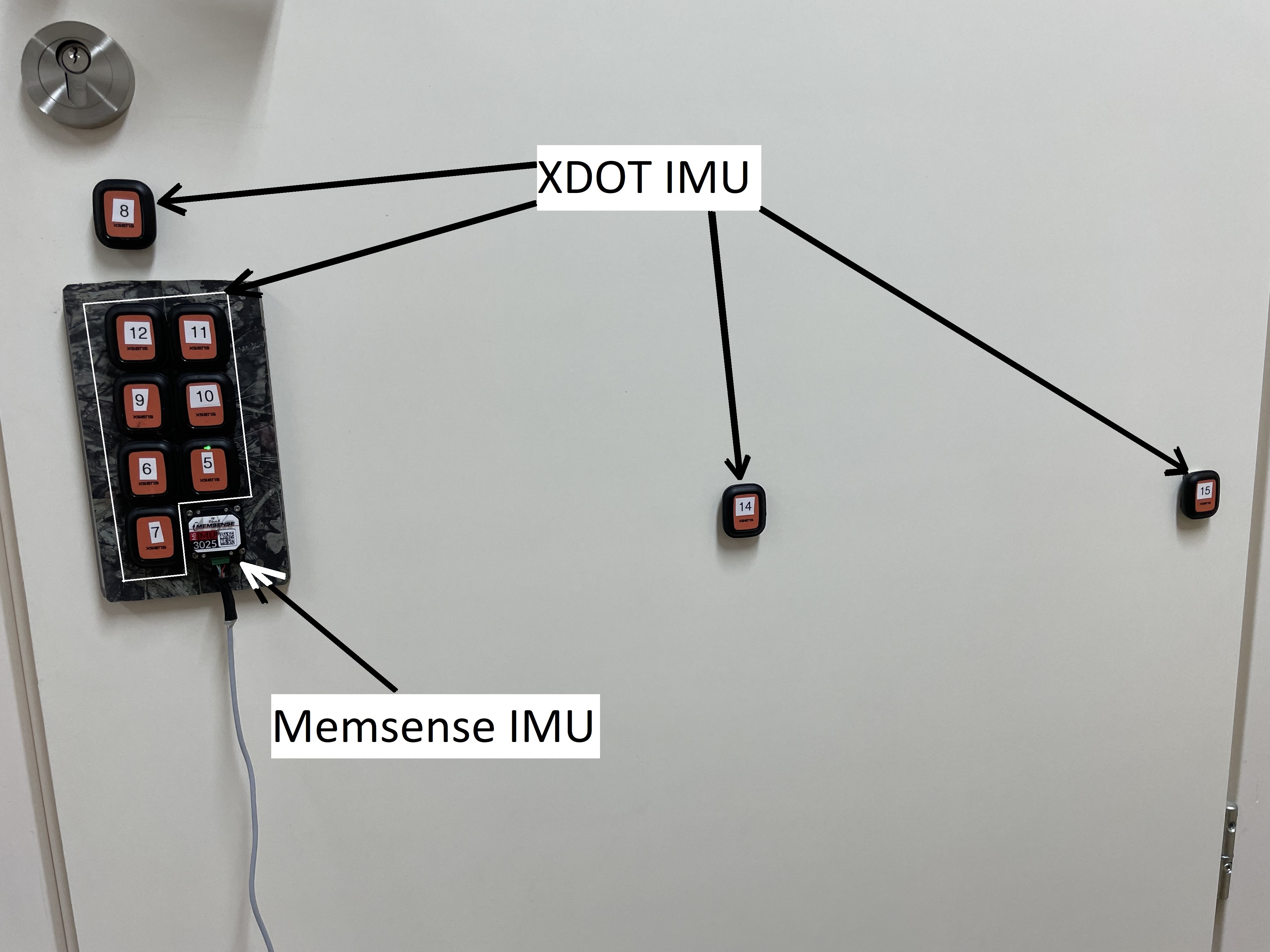}\label{fig:exp_3}}
  \caption{Experimental setups used for dataset generation: DOT IMUs generated raw IMU readings and Memsense IMU generated the ground truth.}
\end{figure}

\noindent
% Three experimental setups were used to generate IMU data, having different number of IMU sensors or position on a door surface. Setup 1 is presented in a Figure \ref{fig:exp_1}: IMU sensors were placed on a door surface next to the handle, 2 different IMU sensors were used, to test the concept of recording IMU data with different sensors simultaneously. Setup 2 is similar to the Setup 1 in number and types of IMU sensors, experiments were performed with three different positions of sensors: next to the door handle, in the middle of the door, and next to the door hinge, to test the influence of sensor position at recorded data and results. Setup 3 used 11 IMU sensors placed on different locations on the door to record as much IMU data as possible.
Our dataset was recorded using two types of IMUs: Memsense MS-IMU3025 \cite{memsenseonline} and Movella Xsens DOT \cite{xdotman}. The Memsense MS-IMU3025 was used to generate the ground-truth (GT) readings. This IMU has a gyroscope bias instability of 0.8°/h over the axis of interest Z and was recorded at 250Hz.  The Movella Xsens DOT IMUs were used as units under test. It has a gyroscope bias instability of 10°/h over the axis of interest Z and was recorded at 120Hz. \\
Three experimental setups were designed and implemented to generate raw inertial data. The setups differ in the number of inertial sensors used and their location on the door surface:
\begin{enumerate}
\item \textbf{Setup 1} Two IMUs—the Memsense and a single DOT—were placed on the door surface next to the handle, as shown in Figure \ref{fig:exp_1}. 
\item \textbf{Setup 2} is similar to Setup 1, but a different DOT sensor was used, and sensors were placed at three different locations consecutively: first next to the door handle, then in the middle of the door, and finally next to the door hinge, to test the influence of sensor position on recorded data and results.
\item \textbf{Setup 3} Consists of the Memsense and ten DOTs placed on a different door (than in Setup 1 and 2) in different positions at the same time. Eight DOTs were placed near the door handle, one in the middle, and one near the door hinge. This setup allowed recording more data in each experiment, and is shown in Figure \ref{fig:exp_3}
\end{enumerate}

% % ----- photo of setup here--------

% We used two types of IMU sensors: up to 10 affordable Xsens DOT IMUs (produced by Movella, formerly Xsens Technologies B.V.) with implicit gyroscope bias instability of 10 \textdegree \slash h (as specified in the user manual on their website \url{https://www.xsens.com/hubfs/Downloads/Manuals/Xsens%20DOT%20User%20Manual.pdf}) with a sampling rate of 120 Hz and high-precision Memsense MS-IMU3025 Inertial Measurement Unit with gyroscope bias instability of 0.8 \textdegree \slash h over the axis of interest Z (as specified on their website \url{https://www.memsense.com/products/ms-imu3025#specifications}) with a sampling rate of 250 Hz.
\noindent
Each recording session comprised several experiments with a duration of 60-120 seconds each. Each experiment consisted of a series (usually 9-11) of openings and closings of the door at a normal speed (like a person would usually open and close the door) to a predetermined angle. Each opening and each closing was followed by a stop (an absence of movement) for 2-5 seconds. After each opening the door was closed shut. \\
Below we elaborate on the specific recording time and setup of the train, validation, and test datasets.
% Ground truth was calculated from high-precision Memsense MS-IMU3025 Inertial Measurement Unit with gyroscope bias instability of 0.8 \textdegree \slash h over the axis of interest Z (as specified on their website \url{https://www.memsense.com/products/ms-imu3025#specifications}). Readings are recorded with a rate of 250 Hz. IMU readings were processed with Madgwick algorithm to calculate Euler angles. Every time the door was closed the Madgwick algorithm was started anew, thus setting sensor drift to zero (explained in section \ref{sec:data_proc_train}).

\subsection{Train and validation datasets}
%1807 Training Process.  you must add the loss function you use and all relevant information about the training process. The date of the recording is irrelevant for the title - use Dataset1, Dataset 2 and etc..
%Alex done - training process in the next session (data processing and model training)
\noindent
The training dataset comprising IMU data recorded in Sessions 1, 2, and 3 is explained below.
\noindent
\subsubsection{\textbf{Session 1}} 
\begin{figure}[ht]
  \centering
  \includegraphics[width=3.5in]{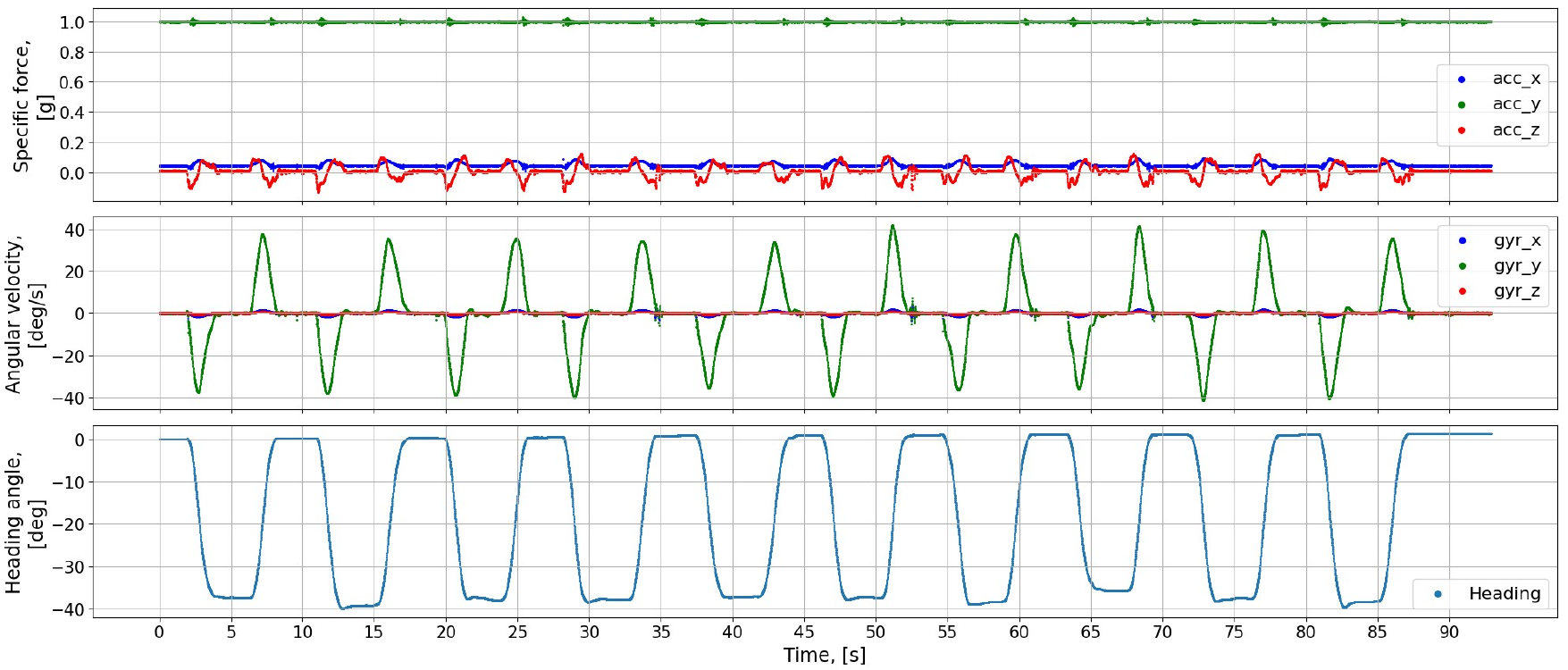}
  \caption{An example of data recorded in Session 1.} \label{fig:session1_example_data}
\end{figure}

The lab door at the Israel Oceanographic and Limnological Research Institute on Tel Shikmona, Haifa, Israel, with experimental Setup 1. \\
% The experimental setup is presented in Figure \ref{fig:exp_1}. One Xsens DOT IMU and one Memsense MS-3025 were placed on the door surface next to the door handle next to each other and aligned that way that the main movement took place on their Z axis.
In this session ten experiments were recorded in total to examine the overall performance and data processing. In each experiment the door was opened to the same angle and closed shut ten times. The approximate angles for the main experiments were in the order of recording: 15\textdegree, 30\textdegree, 45\textdegree, 60\textdegree, 75\textdegree, 90\textdegree. In four test experiments the door was opened and closed three times. The approximate angles for those experiments were 15\textdegree, 45\textdegree, and 90\textdegree. An example of the DOT raw data of the accelerometers and gyroscopes as well as the GT heading is presented in Figure \ref{fig:session1_example_data}.

% Main experiments were later used in building a training dataset. IMU \#14 generated approximately 7.2\% of training / validation data and later was used to test the models on (making it a possible source of data leak).
\noindent
\subsubsection{\textbf{Session 2}} 
\begin{figure}[!ht]
\centering
  \includegraphics[width=3.5in]{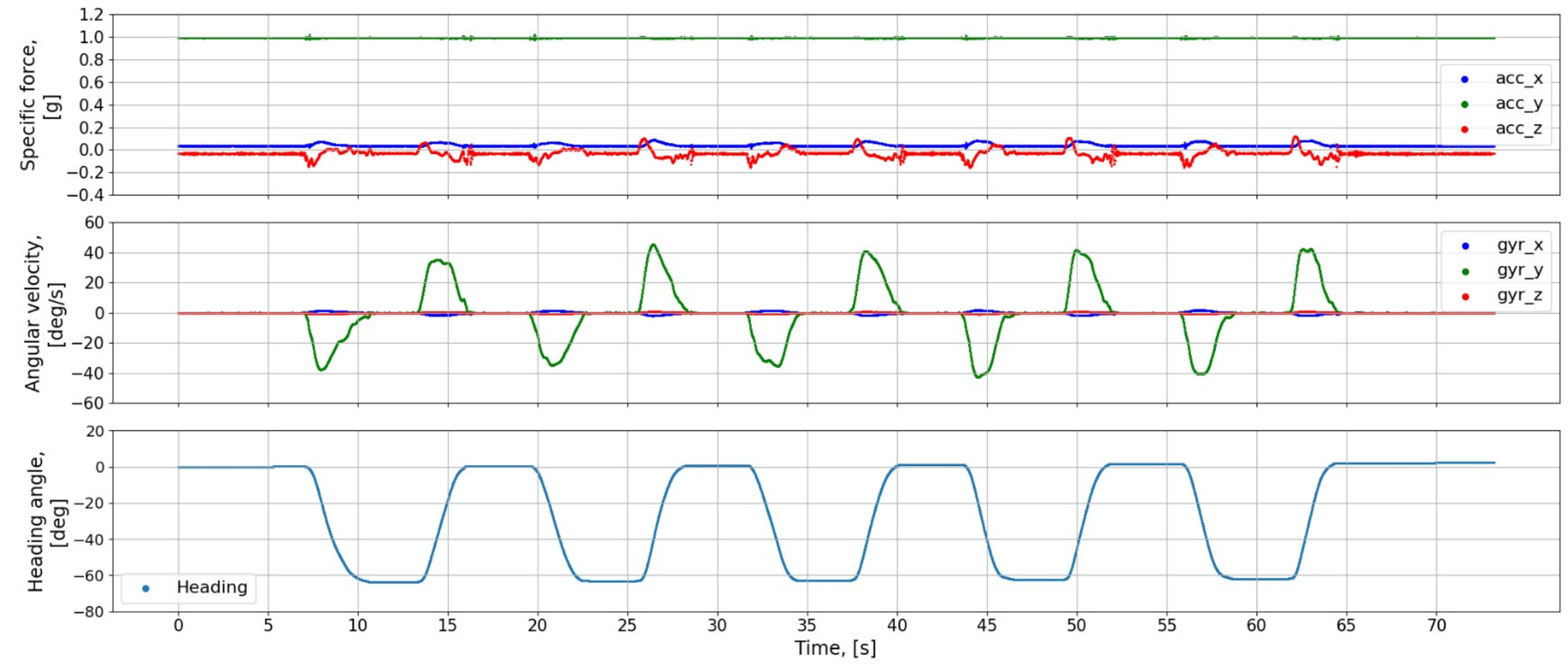}
  \caption{An example of data recorded in Session 2.}\label{fig:session2_example_data}
\end{figure}

The Autonomous Navigation and Sensor Fusion Lab door at the University of Haifa, Israel, with experimental Setup 2. \\
In this session, three experiments were recorded according to the Setup 2. An additional experiment was recorded in a distant location from the door hinge with IMUs rotated 90\textdegree clockwise. At each location two experiments were recorded with a total of eight experiments. In each experiment the door was opened to a predefined angle and closed shut five times. The approximate chosen angles for the experiments were 32\textdegree and 64\textdegree. \\
An example of the DOT raw data of the accelerometers and gyroscopes as well as the GT heading is presented in Figure \ref{fig:session2_example_data}.

% Experiments recorded in the distant position (horizontally oriented) were used in building a training dataset.
\noindent
\subsubsection{\textbf{Session 3}} 
\begin{figure}[!ht]
\centering
  \includegraphics[width=3.5in]{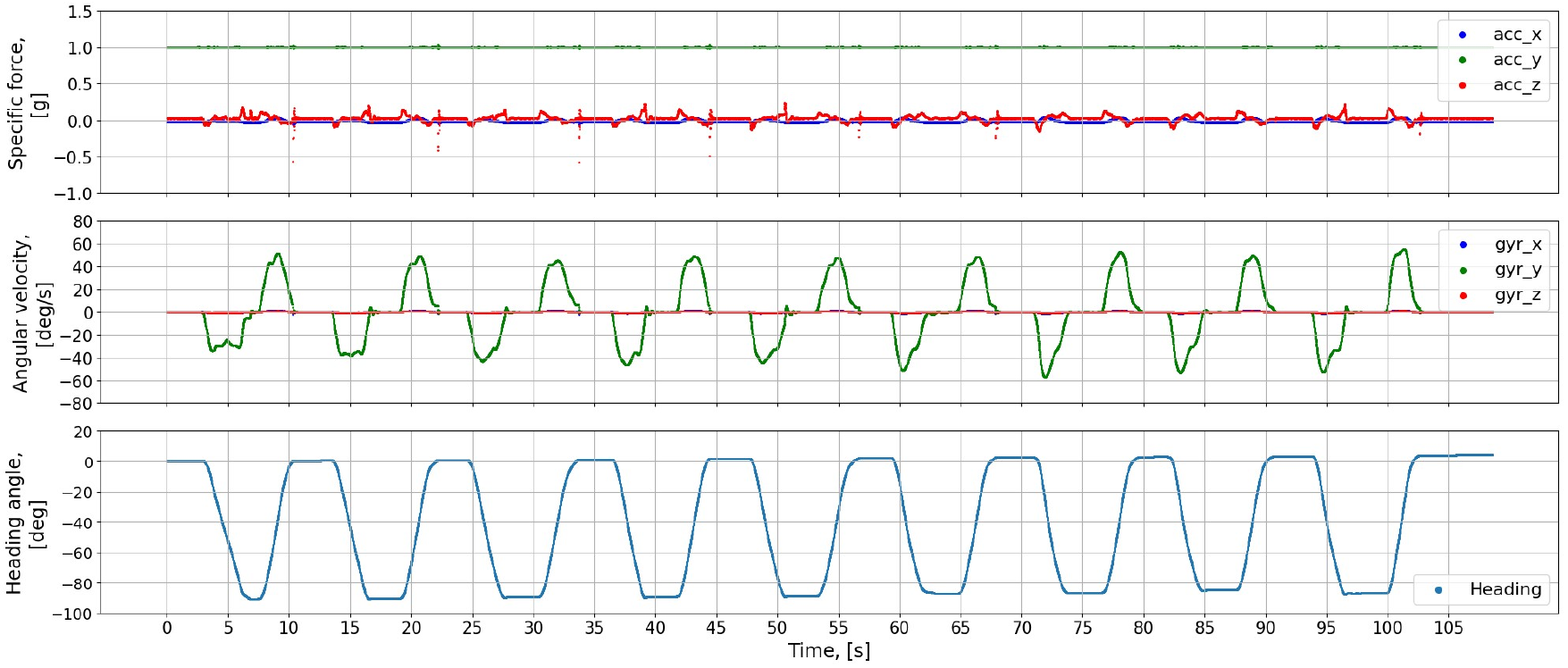}
  \caption{An example of data recorded in Session 3.}\label{fig:session3_example_data}
\end{figure}

The Autonomous Navigation and Sensor Fusion Lab door at the University of Haifa, Israel, with experimental Setup 3. \\
In this session eight experiments were recorded in total. In each, the door was opened to a predefined angle and closed shut 10-11 times. In the last three experiments the door was opened to 90\textdegree~at different speeds: relatively slow, medium, and fast.  \\
An example of the DOT raw data of the accelerometers and gyroscopes as well as the GT heading is presented in Figure \ref{fig:session3_example_data}.

% Most of the recorded data on this session was used in construction of a training dataset. IMU \#5 was used in this session, generating approximately 11.5 \% of training/validation data (possible source of data leak).
% IMU \#12 was never used in training dataset at all, making it a clean source of test data.
% Ground-truth (calculated from Memsense MS-IMU3025 which records with the sampling rate of 250 Hz) was re-sampled to 120 Hz to be consistent with the recorded IMU data from Xsens DOT. Each training sample consists of series of 20 consecutive IMU readings. Training dataset was further split to train and validation dataset (80\% \ 20\%), validation dataset used to look upon validation loss metrics during training. Training dataset consists of 34340 samples and validation dataset consists of 8585 samples.
\vspace{-4mm}
\subsection{Test dataset}
\noindent
The test dataset was recorded last, during Session 3. This was a scenario of using the door for 90 minutes of non-stop recording. The door was opened and closed at will, with several people walking in and out of the internal laboratory room. No specified angle or pattern was used. The door was opened and closed 31 times during this period. \\
\noindent
The GT heading angle was generated using the Memsense IMU3025 IMU readings. It was placed next to the low-cost IMUs that are the units under test.
% The experimental setup consisted of 10 Xsens DOT IMUs and 1 Memsense MS-IMU3025, the photo can be seen on \ref{fig:exp_2}.

\subsection{Dataset summary}
\noindent
The summary of the dataset used for training, validation, and testing is presented in Table \ref{tab:exp_summary}. In total the train dataset has a duration of 95.3 minutes and the validation dataset is 23.8 minutes in length. We used the following notation for the inertial sensor locations:  EP1 = close to the door handle, EP2 = in the middle of the door, and EP3 = close to the door hinge. The test dataset has a duration of 271.9 minutes. It also includes DOT \#12 recordings, which are not present in the train and validation datasets.

% please fill the TBD.
% when you talk about the duration using 10 IMUS, this is the total or for each IMU?
% if this is for each IMU this should be emphasized. 

% The table is great yet requires some revisions.
% in the experiment description please use one of the three: next to the door handle , in the middle of the door , and next to the door hinge .
% do this except when the speed is used in the description (bottom lines).
% what do you mean by no? if this data was not used then remove it from the table
% instaed of length use Duration [s]

\begin{table}[ht]
\caption{Dataset summary showing the description and duration of each experiment. In total, the dataset includes 391 minutes of recorded data.}
\begin{tabular}{ccccc}
\multicolumn{1}{m{0.4cm}}{\centering \textbf{Setup \#}}  & \multicolumn{1}{m{1cm}}{\centering \textbf{Duration, s}} & \multicolumn{1}{m{0.7cm}}{\centering \textbf{\#DOT IMUs}} & \multicolumn{1}{m{2.9cm}}{\centering \textbf{IMU sensors location, Experiment description}} & \multicolumn{1}{m{1cm}}{\centering \textbf{Used in dataset}} \\ \hline
1 & 86.9 & 1 & EP1, 5 angles & train \& val   \\ \hline
% 1 & 91.4 & 1 & Far from hinge & no \\ \hline
1 & 92.9  & 1 & EP1, 6 angles & train \& val  \\ \hline
1 & 106.3 & 1 & EP1, 6 angles  & train \& val  \\ \hline
1 & 113.2 & 1 & EP1, 6 angles & train \& val  \\ \hline
1 & 118.9 & 1 & EP1, 6 angles  & train \& val  \\ \hline
2  & 63.2  & 1 & EP1, 6 angles & train \& val   \\  \hline
2 & 71.5  & 1 & EP1, 6 angles  & train \& val   \\  \hline
% 2 & 66.5 & 1  & Median position  & no  \\  \hline
% 2  & 70.9 & 1 & Median position & no   \\  \hline
% 2  & 60.5 & 1  & Close to the hinge  & no  \\  \hline
% 2  & 71.5  & 1 & Close to the hinge & no  \\  \hline
% 2 & 68.4  & 1 & Far from hinge & no  \\  \hline
% 2 & 67.7 & 1 & Far from hinge & no \\  \hline
3  & 868.1 & 8 & EP1, EP2, EP3, 1 angle & train \& val  \\  \hline
3 & 870.6 & 8 & EP1, EP2, EP3, 1 angle & train \& val   \\  \hline
3 & 766.1 & 8  & EP1, EP2, EP3, 1 angle  & train \& val \\  \hline
3 & 789.4 & 8 & EP1, EP2, EP3, 1 angle & train \& val \\  \hline
3 & 697.4  & 8 & EP1, EP2, EP3, 1 angle & train \& val  \\  \hline
3 & 645.1 & 8 & EP1, EP2, EP3, Fast speed  & train \& val  \\  \hline
3 & 828.4 & 8 & EP1, EP2, EP3, Medium speed & train \& val \\  \hline
3 & 1028.7 & 8  & EP1, EP2, EP3, Slow speed & train \& val \\  \hline
3  & 16312.5  & 3 & EP1, EP2, EP3, Test scenario & test                                                               
\end{tabular}
\label{tab:exp_summary}
\end{table}
% \section{Data processing, model implementation, and training} \label{sec:data_processing}
\subsection{Preprocessing steps: train dataset} \label{sec:data_proc_train}
\begin{figure}[!ht]
\centering
  \includegraphics[width=3.5in]{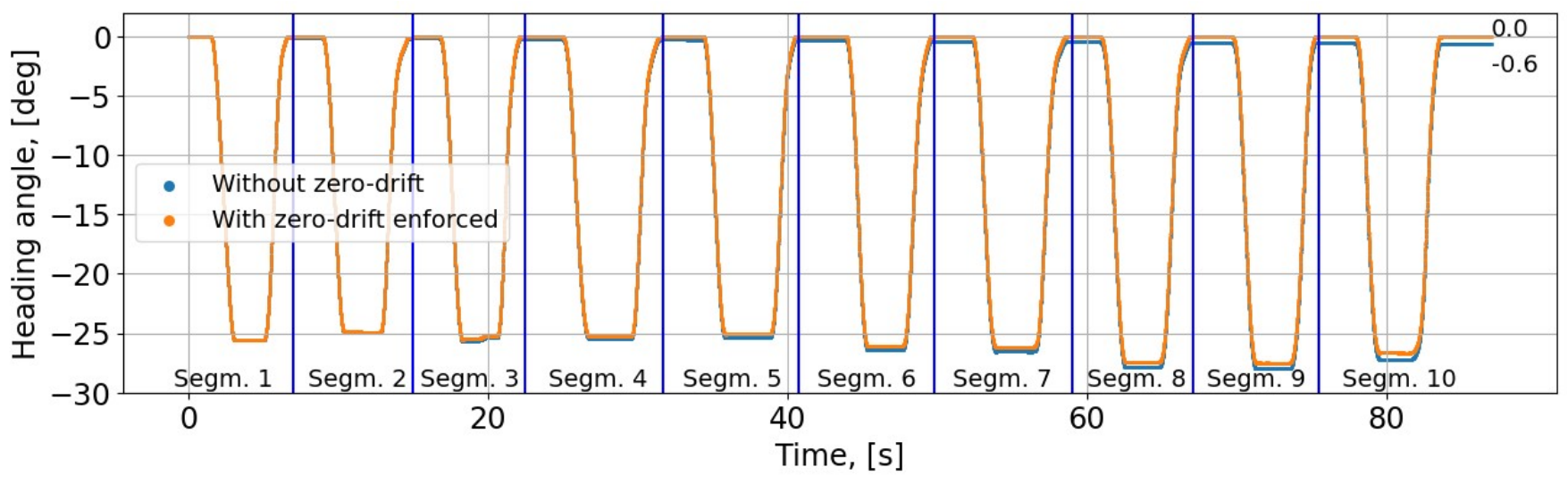}
  \caption{Setting sensor drift to zero for training data. Note the values of the final sensor drift for processed and unprocessed data on the left of the plot.}\label{fig:raw_vs_cut}
\end{figure}

\noindent
To improve the accuracy, two steps were applied to the train and validation datasets. \\
\noindent
1) \textbf{Gyro calibration}:  First order stationary gyro calibration was done for both DOT and Memsense IMUs while the door was shut. The calibration window was 30 to 50 data points. These estimated biases, in each axis, were subtracted from gyroscope readings.\\
\noindent
2) \textbf{Enforcing zero drift}: Zero drift was enforced for each closing of the door for both DOT and Memsense IMUs. As we know that in each experiment the door was shut for two to five seconds between openings, we applied an algorithm that detects periods of the door being shut, so the starting angle is set to zero at the moment the door is shut closed.
\noindent
Figure \ref{fig:raw_vs_cut} shows an example of the Memsense IMU with and without enforcing zero drift. 

% to zero with every closing of the door for both Xsens DOT and Memsense IMUs. Because sensor errors accumulate over time the resulting angle at the end of each experiment is usually different from zero, as depicted on Figure \ref{fig:different_sensor_drift}. We call this effect \textit{"sensor drift"}.

% \begin{figure}[h!]
% \centering
%   \includegraphics[width=3.5in]{pdf/Figure8.pdf}
%   \caption{An example of sensor drift from Memsense IMU (blue) and Xsens DOT IMU (orange).}\label{fig:different_sensor_drift}
% \end{figure}

% In each experiment the door was closed shut for 2-5 seconds between openings so we could know exactly that the desired angle should be zero. A simple algorithm was introduced that detects periods of the door being shut, so the starting angle is set to zero at the moment the door is shut close (Figure \ref{fig:raw_vs_cut}).

\vspace{-4mm}
\subsection{Preprocessing steps: test dataset} \label{sec:data_proc_test}
\begin{figure}[!ht]
\centering
  \includegraphics[width=3.5in]{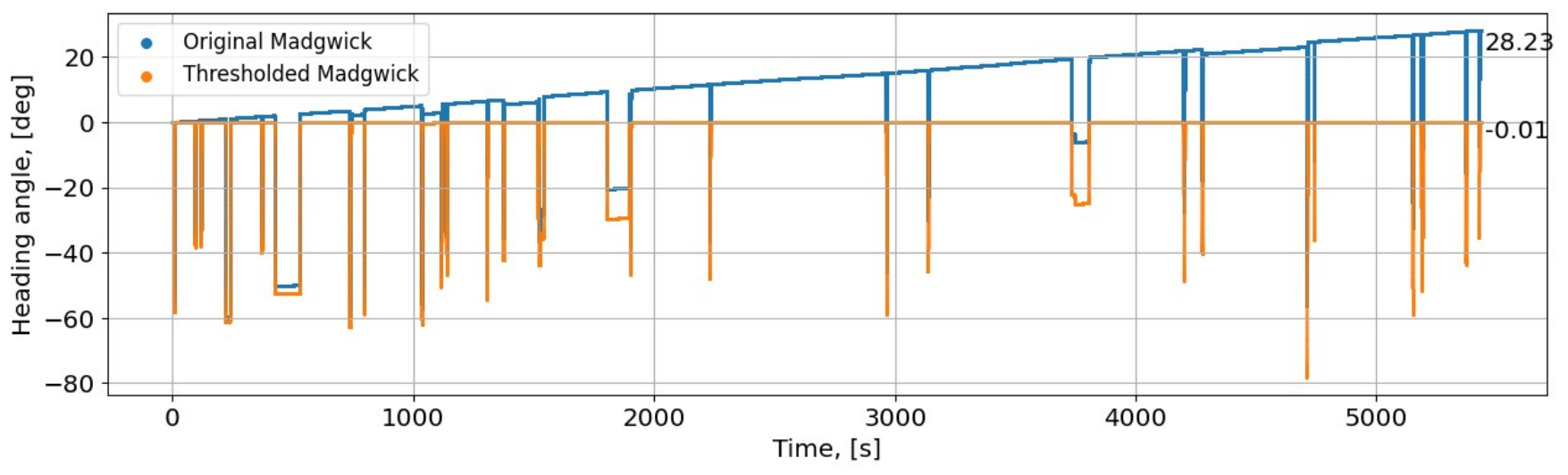}
  \caption{Raw (blue) and threshold-processed GT (orange) heading angles for the test dataset.}\label{fig:raw_vs_thr_gt}
\end{figure}

\noindent
A modified ”thresholded” version of Madgwick algorithm was introduced to calculate GT heading angle from these IMU readings. At each time frame it compares the gyroscope vector norm with a predefined threshold, and, if the vector norm is below the threshold, the algorithm assumes that there is no movement at that specific time frame, and keeps the last value of the heading angle. As a consequence, the GT heading was altered to reflect the actual conditions of the shut door periods as shown in Figure \ref{fig:raw_vs_thr_gt}.

\vspace{-4mm}

\subsection{Model implementation and training}
\begin{figure}[ht]
  \begin{center}
  \includegraphics[width=3.5in]{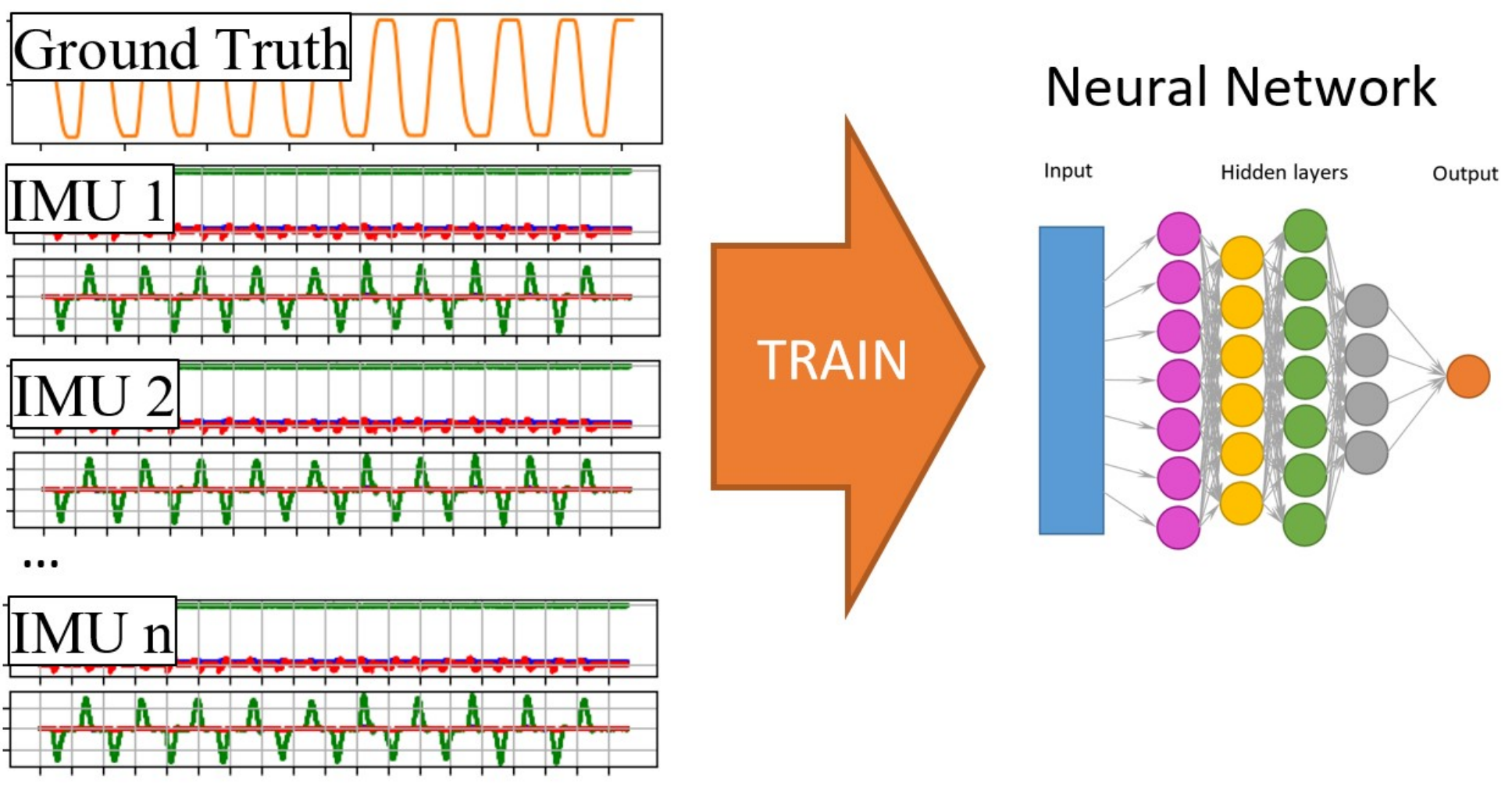}
  \caption{Proposed training approach with heading GT from a higher grade IMU.}\label{fig:prop_training}
  \end{center}
\end{figure}

\noindent
The approach for training includes generating training data from lower-grade  IMUs, recorded simultaneously, with the heading GT from a higher grade IMU as illustrated in Figure \ref{fig:prop_training}. In this setting the deep-learning model is able to generalize over different sensor parameters and behavior.

\begin{figure}[!ht]
\centering
  \includegraphics[width=3.5in]{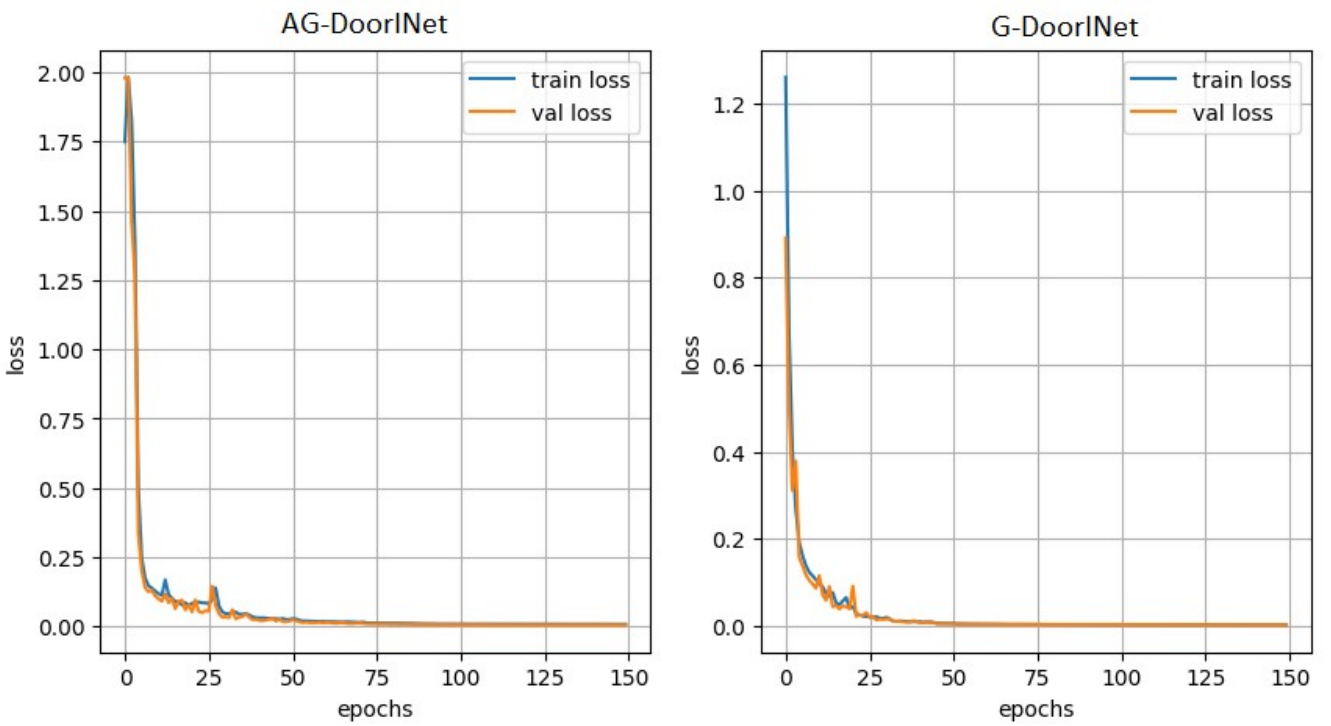}
  \caption{Train and validation losses for AG-DoorINet (left) and G-DoorINet (right).}\label{fig:train_loss}
\end{figure}

\noindent
Model-based algorithms were implemented in the Python programming language; the Madgwick algorithm was already implemented in \texttt{imufusion} Python package (\url{https://github.com/xioTechnologies/Fusion}). All our modifications of Madgwick algorithm were implemented as additions to the implementation mentioned above. \\
Our DoorINet deep-learning models were implemented in the PyTorch framework version 2.1.0+cu116 and were trained on the train dataset using a validation dataset for monitoring the training process using a back-propagation algorithm. Weights of our networks were initialized using the Xavier uniform initialization scheme \cite{xavier_init_paper}. The hardware used for training was an ASUS TUF Gaming A17 laptop with AMD Ryzen 7 4800H CPU, 16GB RAM, and Nvidia GeForce GTX 1660 Ti GPU. We used a Huber loss function (designed for regression problems) with default parameters, described as

\begin{equation}\label{eq:huber_loss1}
    HuberLoss(\boldsymbol{\mathrm{Y}},\boldsymbol{\mathrm{\hat{Y}}}) = mean(l_1, ... , l_n)
\end{equation}
where $\boldsymbol{\mathrm{Y}}$ is the ground-truth vector, $\boldsymbol{\mathrm{\hat{Y}}}$ is the model output vector, $n$ is the number of model outputs, and $l_i$ is calculated as a function of $i$-th elements $y_i$ of the ground truth and $\hat{y}_i$ of the model output by
\begin{equation}\label{eq:huber_loss2}
    l_i = 
\begin{cases}
    0.5(y_i - \hat{y}_i)^2, & \text{if }  |y_i - \hat{y}_i| < 1, \\
    (|y_i - \hat{y}_i| - 0.5), & \text{otherwise}.
\end{cases}
\end{equation}
\noindent
Training was performed over 150 epochs using the \texttt{AdamW} optimizer and \texttt{ReduceOnPlateau} scheduler with an initial learning rate of 1e-3, reducing the learning rate by two after waiting for three epochs of the loss value dropping less than 0.01. The history of loss values during the training phase are presented in Figure \ref{fig:train_loss}.

\noindent
For our results to be as reproducible as possible we introduced seed value as a hyperparameter for the training procedure. The seed value initializes the pseudorandom number generator in \texttt{PyTorch} and \texttt{NumPy} libraries, so every time a generator is used to generate a number (i.e., during the initialization phase), it does so in a deterministic fashion. Therefore, a training optimization process starts from a fixed point on a loss hypersurface, ensuring convergence to the same values on every run.\\
We performed training with ten randomly chosen seed values. Results presented in this paper were obtained with seed values that produce results close to the average, which were seed value 700 for AG-DoorINet and 35 for G-DoorINet.
%\vspace{-4mm}
\section{Experimental results} \label{sec:results}

\subsection{Metrics}
\noindent
Several metrics were used to measure performance of different models:

\begin{itemize}
    % \item   Mean Squared Error is a sum of squares of differences between the processed \textit{"thresholded"} ground-truth recording and the same recording reconstructed from model outputs:

    % \begin{equation}\label{eq:mse_formula}
    %     MSE = \frac{1}{n} \sum_{i=1}^n (Y_i - \hat{Y}_i) 
    % \end{equation}
    % where $n$ is a number of model outputs in a test dataset, $Y_i$ denotes the ground-truth (real) $i$-th sample and  $\hat{Y}_i$ is the $i$-th value predicted by the model;

    \item \textbf{Root mean squared error (RMSE)} is the square root of the sum of squares of difference between the processed heading GT, $\boldsymbol{\mathrm{Y}}$, and the estimated heading, $\boldsymbol{\mathrm{\hat{Y}}}$:

    \begin{equation}\label{eq:rmse_formula}
        RMSE = \sqrt{\frac{1}{n} \sum_{i=1}^n (y_i - \hat{y}_i)} 
    \end{equation}
    where $n$ is a number of model outputs.

    \item  \textbf{Last point difference (LPD)} between the absolute value of the  difference of the sum of GT heading to its estimate:
    
    \begin{equation}\label{eq:aad_formula}
        LPD = \bigg |\sum_{i=1}^n y_i - \sum_{i=1}^n \hat{y}_i \bigg |.
    \end{equation}

    \item  \textbf{Maximum absolute difference (MAD)} is defined as the maximum value of the LPD over all instances during the evaluation: 

    \begin{equation}\label{eq:mad_formula}
        MAD = \max_{i} \bigg\{ \bigg |\sum_{i=1}^n y_i - \sum_{i=1}^n \hat{y}_i \bigg | \bigg\}.
    \end{equation}
    
\end{itemize}

% LGC-Net architecture includes special kind of convolution layers called depthwise separable convolution \cite{chollet2017xception} which consists of depthwise convolution, batch normalization, GELU activation, pointwise convolution and dropout layers; it also includes special kind of attention mechanism called Large Kernel Attention \cite{guo2022visual}, which consists of depthwise convolution, depthwise dilation convolution and 1x1 convolution.
% LGC-Net performs denoising of IMU readings of cheap MEMS-based inertial sensors. Raw 3-axis accelerometer and 3-axis gyroscope readings are inputs to the network which outputs correction parameters for inertial readings. The attitude is then calculated by integrating of the noise-free gyroscope outputs.
% We were unable to acquire neither the trained LGC-Net model nor the source code for it, and so we had to implement the proposed approach ourselves. We implemented the network architecture and trained it on our data. Unfortunately we were not able to implement the loss function proposed by the authors, so we used Huber loss function \eqref{eq:huber_loss1}.
\vspace{-1mm}
\subsection{Results}
% Results are presented for all testing data in the table \ref{tab:results} and individually for each IMU on Figures \ref{fig:res_imu12}-\ref{fig:res_imu5}.
\noindent
As described in Table \ref{tab:exp_summary}, the test dataset includes recordings of DOT \#12 that were not used in the train and validation dataset, IMU \#14 that was used in \textbf{Session 1} to generate 7.2\% of the data used for training and validation, and IMU \#5 that was used in \textbf{Session 3} to generate 11.5\% of the training/validation data sets.

\begin{figure*}
\centering
  \includegraphics[width=7in]{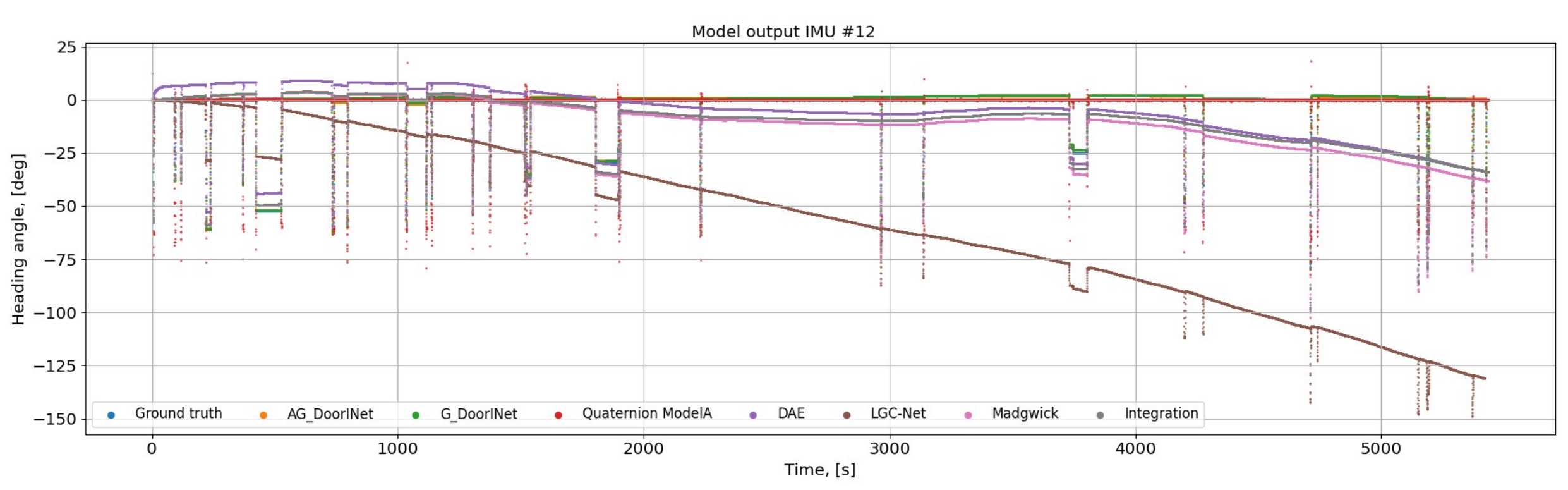}
  \caption{Estimated heading angle obtained by seven different approaches and heading GT for the test dataset containing IMU \#12 readings.}\label{fig:res_imu12}
\end{figure*}

\begin{figure*}
\centering
  \includegraphics[width=7in]{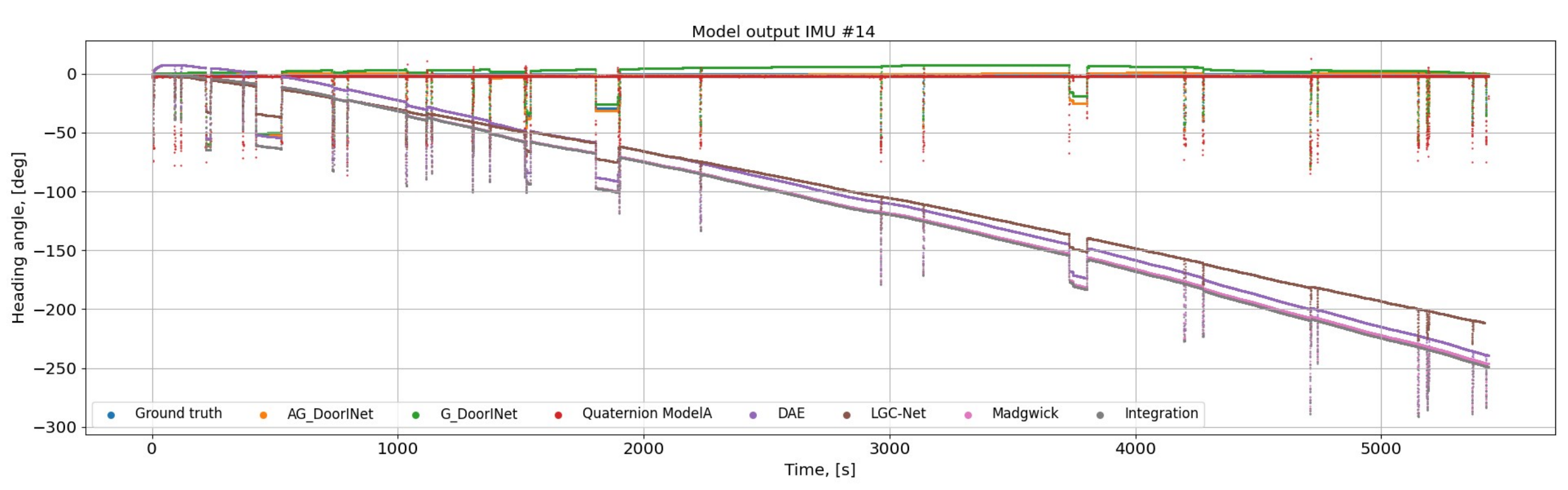}
  \caption{Estimated heading angle obtained by seven different approaches and heading GT for the test dataset containing IMU \#14 readings.}\label{fig:res_imu14}
\end{figure*}

\begin{figure*}
\centering
  \includegraphics[width=7in]{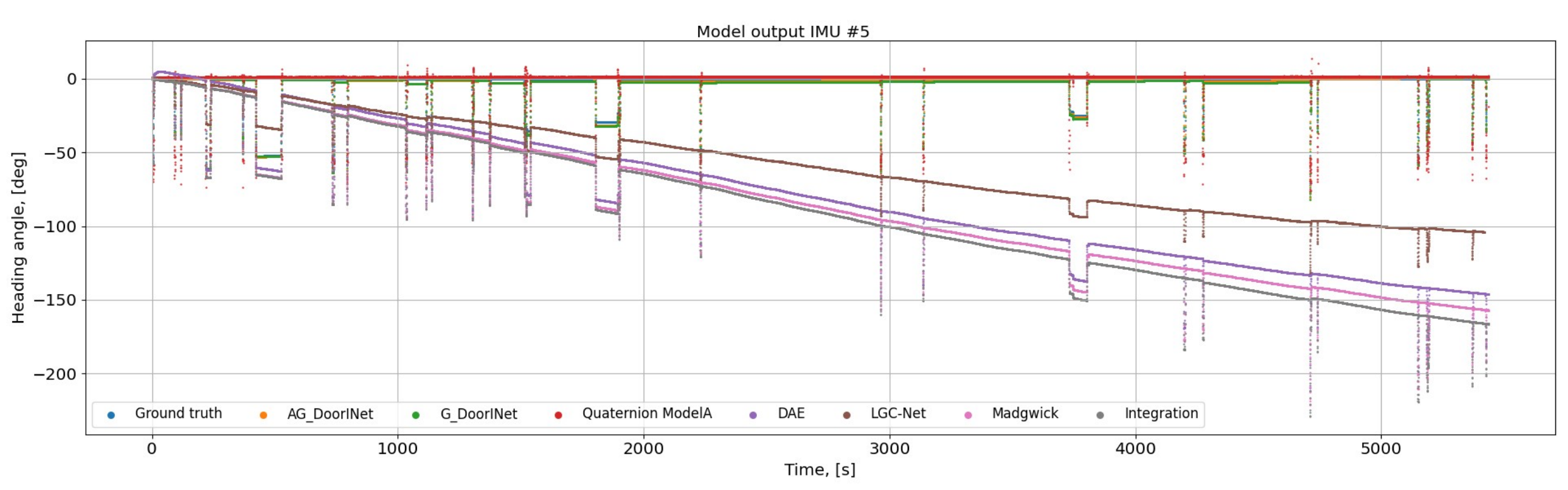}
  \caption{Estimated heading angle obtained by seven different approaches and heading GT for the test dataset containing IMU \#5 readings.}\label{fig:res_imu5}
\end{figure*}

% Seven model-based and data-driven algorithms were compared on the recorded real-life scenario (test data):

% \begin{itemize}
%     \item \textit{Model-based:} Madgwick MARG algorithm (baseline) (see section \ref{sec:madgwick_alg})
%     \item \textit{Model-based:} Gyroscope integration over time (see section \ref{sec:gyro_form})
%     \item \textit{Hybrid approach:} DAE model (from \cite{EranDAE})
%     \item \textit{Data-driven:} Quaternion Model A (from \cite{ASGHARPOORGOLROUDBARI2023113105})
%     \item \textit{Data-driven:} LGC-Net (our interpretation) (from \cite{lgcnet})
%     \item \textit{Data-driven:} DoorINet model (ours) with 6-axis input (accelerometer and gyroscope) (see Table \ref{tab:6axis_structure})
%     \item \textit{Data-driven:} DoorINet (our) model with 3-axis input (gyroscope only) (see Table \ref{tab:3axis_structure})
% \end{itemize} 
\noindent
Figures \ref{fig:res_imu12}-\ref{fig:res_imu5} present the heading estimation versus time for three different IMUs applied on the test dataset. The heading estimation approaches include our two DoorINet architectures and five other approaches as described in Section \ref{sec:problem_formulation}. \\
\noindent
% performance of selected model-based AHRS algorithms (direct gyroscope integration and more complex Madgwick algorithm) and data-driven models: Quaternion Model A, LGC-Net, DAE, AG-DoorINet (our) and G-DoorINet (our) algorithms on IMU data, recorded by DOT IMU \#12, on a real-life scenario over 90 minutes length. GT data was generated by a high-precision Memsense MS-IMU3025. Data generated by IMU \#12 was never used for training of any of DoorINet (our) models. 
As presented in the Figure \ref{fig:res_imu12}, all of the tested algorithms perform worse than the proposed DoorINet models, and their performance becomes worse with time (the difference becomes bigger). The figure also shows that the Quaternion ModelA fails to calculate the correct angle for the door opening. Both of our approaches obtained the best performance where AG-DoorINet has a minimum RMSE of 1.24 degrees. \\
\noindent
In the same manner, Figure \ref{fig:res_imu14} presents the results of the test data recorded with the DOT IMU \#14. This IMU was previously used in recording \textbf{Session 1}, generating 7.2\% of the training and validation data. The Madgwick algorithm is comparable to the gyroscope integration approach. Both of our models perform better than others, where AG-DoorINet achieves the minimum RMSE of 1.03 degrees. \\
\noindent
Figure \ref{fig:res_imu5} presents the performance of all the test data recorded by DOT IMU \#5. This IMU was  used in \textbf{Session 3}, and  generated 11.5\% of all the training and validation data. The DAE hybrid approach improved the result given by gyroscope integration and Madgwick algorithms, but the DoorINet model has better results and lower RMSE values (as low as 1.42 degrees for AG-DoorINet).
\vspace{-4mm}
\subsection{Summary}
\noindent
A summary of all the results described above is presented in Table \ref{tab:results}. 
\begin{table*}
\caption{Summary of the results obtained using three different test datasets.}
\centering
% \begin{adjustbox}{width=\columnwidth}
% \renewcommand{\arraystretch}{2}
\begin{tabular}[b!]{m{8cm}m{3cm}m{3cm}m{3cm}} 
\multicolumn{1}{m{8cm}}{\centering \textbf{Model}}  & 
\multicolumn{1}{m{3cm}}{\centering \textbf{RMSE [deg]}}   & 
\multicolumn{1}{m{3cm}}{\centering \textbf{LPD [deg]}} & 
\multicolumn{1}{m{3cm}} {\centering \textbf{MAD [deg]}} \\
% &  \multicolumn{1}{m{3cm}}{\centering \textbf{Improvement over baseline (RMSE), \%}} \\ 
 & & & \\ \hline

\multicolumn{4}{c}{DOT IMU \#12 (never used for producing training data)} \\ \hline
% \multicolumn{1}{c}{Madgwick MAG algorithm (baseline)}   & \multicolumn{1}{c}{0.64}  & \multicolumn{1}{c}{0.15}    & \multicolumn{1}{c}{58.81} & \multicolumn{1}{c}{---} \\ \hline
\multicolumn{1}{c}{Madgwick algorithm}   & \multicolumn{1}{c}{14.25}  & \multicolumn{1}{c}{38.09}    & \multicolumn{1}{c}{40.47} 
% & \multicolumn{1}{c}{---} 
\\ \hline
\multicolumn{1}{c}{Gyroscope integration}    & \multicolumn{1}{c}{12.23}  & \multicolumn{1}{c}{34.00} & \multicolumn{1}{c}{36.68}
% & \multicolumn{1}{c}{no improvement}   
\\ \hline
\multicolumn{1}{c}{DAE}   & \multicolumn{1}{c}{11.78}  & \multicolumn{1}{c}{38.09}   & \multicolumn{1}{c}{40.47}  
% & \multicolumn{1}{c}{no improvement} 
\\ \hline
\multicolumn{1}{c}{LGC-Net (our implementation)}   & \multicolumn{1}{c}{67.34}  & \multicolumn{1}{c}{130.87} & \multicolumn{1}{c}{130.91} 
% & \multicolumn{1}{c}{no improvement}   
\\ \hline
\multicolumn{1}{c}{Quaternion Model A}   & \multicolumn{1}{c}{10.72}  & \multicolumn{1}{c}{19.67}   & \multicolumn{1}{c}{65.48}   
% & \multicolumn{1}{c}{no improvement} 
\\ \hline
\multicolumn{1}{c}{AG-DoorINet (ours)} & \multicolumn{1}{c}{1.24} & \multicolumn{1}{c}{0.02}  & \multicolumn{1}{c}{7.14} 
% & \multicolumn{1}{c}{???}   
\\ \hline
\multicolumn{1}{c}{G-DoorINet (ours)}  & \multicolumn{1}{c}{1.35} & \multicolumn{1}{c}{0.02}  & \multicolumn{1}{c}{7.25} 
% & \multicolumn{1}{c}{???}  
\\ \hline
\multicolumn{4}{c}{DOT IMU \#14 (used to produce 7.2\% of training data)}      \\ \hline
% \multicolumn{1}{c}{Madgwick MAG algorithm (baseline)}   & \multicolumn{1}{c}{0.57}  & \multicolumn{1}{c}{0.05}  & \multicolumn{1}{c}{39.66} & \multicolumn{1}{c}{---} \\ \hline
\multicolumn{1}{c}{Madgwick algorithm}   & \multicolumn{1}{c}{132.28}  & \multicolumn{1}{c}{246.34}  & \multicolumn{1}{c}{247.87} 
% & \multicolumn{1}{c}{---} 
\\ \hline
\multicolumn{1}{c}{Gyroscope integration}  & \multicolumn{1}{c}{133.65}  & \multicolumn{1}{c}{248.89}  & \multicolumn{1}{c}{250.69} 
% & \multicolumn{1}{c}{no improvement} 
\\ \hline
\multicolumn{1}{c}{DAE}   & \multicolumn{1}{c}{125.83}  & \multicolumn{1}{c}{239.22}   & \multicolumn{1}{c}{240.74}   
% & \multicolumn{1}{c}{no improvement} 
\\ \hline
\multicolumn{1}{c}{LGC-Net (our implementation)}   & \multicolumn{1}{c}{115.48}  & \multicolumn{1}{c}{211.47}     & \multicolumn{1}{c}{211.47} 
% & \multicolumn{1}{c}{no improvement} 
\\ \hline
\multicolumn{1}{c}{Quaternion Model A}  & \multicolumn{1}{c}{10.44}  & \multicolumn{1}{c}{18.81}  & \multicolumn{1}{c}{67.76} 
% & \multicolumn{1}{c}{no improvement} 
\\ \hline
\multicolumn{1}{c}{AG-DoorINet (our)} & \multicolumn{1}{c}{1.03}  & \multicolumn{1}{c}{0.02}  & \multicolumn{1}{c}{6.26} 
% & \multicolumn{1}{c}{???}   
\\ \hline
\multicolumn{1}{c}{G-DoorINet (our)}  & \multicolumn{1}{c}{4.62}  & \multicolumn{1}{c}{0.02}   & \multicolumn{1}{c}{9.16} 
% & \multicolumn{1}{c}{???}   
\\ \hline

\multicolumn{4}{c}{DOT IMU \#5 (used to produce 11.5\% of training data)}    \\ \hline
% \multicolumn{1}{c}{Madgwick MAG algorithm (baseline)}   & \multicolumn{1}{c}{0.46}   & \multicolumn{1}{c}{0.13}  & \multicolumn{1}{c}{19.54} & \multicolumn{1}{c}{---}   \\ \hline
\multicolumn{1}{c}{Madgwick algorithm}   & \multicolumn{1}{c}{96.05}   & \multicolumn{1}{c}{156.96}  & \multicolumn{1}{c}{160.89} 
% & \multicolumn{1}{c}{---}   
\\ \hline
\multicolumn{1}{c}{Gyroscope integration}  & \multicolumn{1}{c}{100.78}  & \multicolumn{1}{c}{166.33} & \multicolumn{1}{c}{170.48}
% & \multicolumn{1}{c}{no improvement}    
\\ \hline
\multicolumn{1}{c}{DAE}   & \multicolumn{1}{c}{89.62}  & \multicolumn{1}{c}{146.09}   & \multicolumn{1}{c}{150.12}   
% & \multicolumn{1}{c}{no improvement} 
\\ \hline
\multicolumn{1}{c}{LGC-Net (our implementation)}  & \multicolumn{1}{c}{65.20}  & \multicolumn{1}{c}{103.87}  & \multicolumn{1}{c}{104.01} 
% & \multicolumn{1}{c}{no improvement}   
\\ \hline
\multicolumn{1}{c}{Quaternion Model A} & \multicolumn{1}{c}{10.84}  & \multicolumn{1}{c}{18.92}    & \multicolumn{1}{c}{64.34}
% & \multicolumn{1}{c}{no improvement}    
\\ \hline
\multicolumn{1}{c}{AG-DoorINet (our)} & \multicolumn{1}{c}{1.42} & \multicolumn{1}{c}{0.02} & \multicolumn{1}{c}{11.46} 
% & \multicolumn{1}{c}{???} 
\\ \hline
\multicolumn{1}{c}{G-DoorINet (our)}  & \multicolumn{1}{c}{1.93} & \multicolumn{1}{c}{0.02} & \multicolumn{1}{c}{12.45} 
% & \multicolumn{1}{c}{???}  
\\ \hline
\end{tabular}
% \end{adjustbox}
\label{tab:results}
\end{table*}
% \vspace{-4mm}
\noindent
All the tested algorithms operate only with accelerometer and gyroscope readings. Potentially, the Madgwick algorithm could improve its performance by incorporating magnetic readings, but that would require additional fine tuning and may not be suitable in environments with strong and changing magnetic interference. \\
Our accelerometer and gyroscope AG-DoorINet model shows better performance than the gyroscope-only G-DoorINet on all the metrics and test data, which indicates that the accelerometes measurements contribute to the heading accuracy. When the door is used in an ordinary fashion, acceleration naturally takes place as a human cannot open or close the door without accelerating it. Therefore, this additional information about the movement embedded in the accelerometer readings helps in improving the heading angle calculations. \\
Although all learning approaches were trained on the same training set, ours performed better as it was designed specifically for door dynamics, which is limited compared to the other approaches designed to handle general attitude and heading setups. 
% These results  aligns with the model-based approaches as accelerometers are used to update only Euler roll and pitch angles with.
% Our data-driven DoorINet algorithms may also be susceptible to data leak, as values of LPD and MAD for all the test DOT IMUs that have been used in producing training and validation data are lower than for the DOT IMU \#12 that was ultimately new to all the models.

\section{Conclusion} \label{sec:conclusions}
\noindent
In this work we derived and presented DoorINet; a deep-learning, end-to-end framework to estimate the angle of an opening door using only low-cost inertial sensors. We introduced two models: AG-DoorINet that uses accelerometer and gyroscope IMU readings, and G-DoorINet that uses only gyroscope readings. To evaluate our approaches we recorded a unique dataset using ten low-cost sensors and a corresponding GT heading angle obtained using a higher grade sensor. The train and validation dataset consists of 119 minutes of IMU recordings while the test dataset has 272 minutes of IMU recordings. Included in the test set is a real-world scenario of over 90 minutes duration of a non-stop recording of opening/closing a door. \\
We demonstrated the strength of our proposed approach by comparing it with five model- and learning-based approaches. Our AG-DoorINet approach obtained the best performance (RMSE, MAD, and LPD) across all examined scenarios. Our deep-learning framework was able to generalize over different error parameters of different inertial sensors, and maintained performance for a long duration of 90 minutes. \\
G-DoorINet also outperformed all other approaches yet obtained lower performance compared to AG-DoorINet. Thus, the accelerometer measurements contribute in improving the overall performance, providing additional information about the door motion. \\ 
% Although was worthy to examine, this conclusion fits the well-known theory that accelerometers are used only to update only the roll and pitch angels and have no influence on the heading angle. \\
DoorINet is suitable for different smart home and office applications; for example, smart office, home office, or building management. In addition, DoorINet can be implemented on any other AHRS application. In future work, we aim to extend DoorINet to include estimation of the roll and pitch angles.

\section*{Acknowledgment}
\noindent
Aleksei Zakharchenko is supported by the Maurice Hatter Foundation.

\bibliographystyle{IEEEtran}
\bibliography{IEEEabrv,DoorINet_Article}

\end{document}